\pdfoutput=1
\documentclass[11pt,twoside,a4paper,pdftex,cmspaper,final,collab]{cms-tdr}

\begin{document}\cmsNoteHeader{CFT-09-016}
%%%%%%%%%%%%%%%%%%%%%%%%%%%%%%%%%%%%%%%%%%%%%%%%%%%%%%%%%%%%%%%%%%%%
%
%  Common definitions
%
%  N.B. use of \providecommand rather than \newcommand means
%       that a definition is ignored if already specified
%
%                                              L. Taylor 18 Feb 2005
%%%%%%%%%%%%%%%%%%%%%%%%%%%%%%%%%%%%%%%%%%%%%%%%%%%%%%%%%%%%%%%%%%%%

%%%%%%%%%%%%%%%%%%%%%%%%%%%%%%%%%%%%%%%%%%%%%%%%%%%%%%%%%%%%%%%%%%%%
%
% Hyphenations (only need to add here if you get a nasty word break)
%
\hyphenation{env-iron-men-tal}%    just an example
\hyphenation{had-ron-i-za-tion}
\hyphenation{cal-or-i-me-ter}
\hyphenation{de-vices}
%
% Hyphenations-end
%
% CVS info. These are modified by cvs at checkout time.
% The last version of these macros found before the maketitle will be the one on the front page,
% so only the main file is tracked.
% Edit by hand with care!
\RCS$Revision: 1.32 $
\RCS$Date: 2010/02/05 03:20:33 $
\RCS$Name:  $
%%%%%%%%%%%%% ptdr definitions %%%%%%%%%%%%%%%%%%%%%
%%%%%%%%%%%%%%%%%%%%%%%%%%%%%%%%%%%%%%%%%%%%%%%%%%%%%%%%%%%%%%%%%%%%
%
%  Common definitions
%
%  N.B. use of \providecommand rather than \newcommand means
%       that a definition is ignored if already specified
%
%                                              L. Taylor 18 Feb 2005
%%%%%%%%%%%%%%%%%%%%%%%%%%%%%%%%%%%%%%%%%%%%%%%%%%%%%%%%%%%%%%%%%%%%

% Some shorthand
% turn off italics
\newcommand {\etal}{\mbox{et al.}\xspace} %et al. - no preceding comma
\newcommand {\ie}{\mbox{i.e.}\xspace}     %i.e.
\newcommand {\eg}{\mbox{e.g.}\xspace}     %e.g.
\newcommand {\etc}{\mbox{etc.}\xspace}     %etc.
\newcommand {\vs}{\mbox{\sl vs.}\xspace}      %vs.
\newcommand {\mdash}{\ensuremath{\mathrm{-}}} % for use within formulas

% some terms whose definition we may change
\newcommand {\Lone}{Level-1\xspace} % Level-1 or L1 ?
\newcommand {\Ltwo}{Level-2\xspace}
\newcommand {\Lthree}{Level-3\xspace}

% Some software programs (alphabetized)
\providecommand{\ACERMC} {\textsc{AcerMC}\xspace}
\providecommand{\ALPGEN} {{\textsc{alpgen}}\xspace}
\providecommand{\CHARYBDIS} {{\textsc{charybdis}}\xspace}
\providecommand{\CMKIN} {\textsc{cmkin}\xspace}
\providecommand{\CMSIM} {{\textsc{cmsim}}\xspace}
\providecommand{\CMSSW} {{\textsc{cmssw}}\xspace}
\providecommand{\COBRA} {{\textsc{cobra}}\xspace}
\providecommand{\COCOA} {{\textsc{cocoa}}\xspace}
\providecommand{\COMPHEP} {\textsc{CompHEP}\xspace}
\providecommand{\EVTGEN} {{\textsc{evtgen}}\xspace}
\providecommand{\FAMOS} {{\textsc{famos}}\xspace}
\providecommand{\GARCON} {\textsc{garcon}\xspace}
\providecommand{\GARFIELD} {{\textsc{garfield}}\xspace}
\providecommand{\GEANE} {{\textsc{geane}}\xspace}
\providecommand{\GEANTfour} {{\textsc{geant4}}\xspace}
\providecommand{\GEANTthree} {{\textsc{geant3}}\xspace}
\providecommand{\GEANT} {{\textsc{geant}}\xspace}
\providecommand{\HDECAY} {\textsc{hdecay}\xspace}
\providecommand{\HERWIG} {{\textsc{herwig}}\xspace}
\providecommand{\HIGLU} {{\textsc{higlu}}\xspace}
\providecommand{\HIJING} {{\textsc{hijing}}\xspace}
\providecommand{\IGUANA} {\textsc{iguana}\xspace}
\providecommand{\ISAJET} {{\textsc{isajet}}\xspace}
\providecommand{\ISAPYTHIA} {{\textsc{isapythia}}\xspace}
\providecommand{\ISASUGRA} {{\textsc{isasugra}}\xspace}
\providecommand{\ISASUSY} {{\textsc{isasusy}}\xspace}
\providecommand{\ISAWIG} {{\textsc{isawig}}\xspace}
\providecommand{\MADGRAPH} {\textsc{MadGraph}\xspace}
\providecommand{\MCATNLO} {\textsc{mc@nlo}\xspace}
\providecommand{\MCFM} {\textsc{mcfm}\xspace}
\providecommand{\MILLEPEDE} {{\textsc{millepede}}\xspace}
\providecommand{\ORCA} {{\textsc{orca}}\xspace}
\providecommand{\OSCAR} {{\textsc{oscar}}\xspace}
\providecommand{\PHOTOS} {\textsc{photos}\xspace}
\providecommand{\PROSPINO} {\textsc{prospino}\xspace}
\providecommand{\PYTHIA} {{\textsc{pythia}}\xspace}
\providecommand{\SHERPA} {{\textsc{sherpa}}\xspace}
\providecommand{\TAUOLA} {\textsc{tauola}\xspace}
\providecommand{\TOPREX} {\textsc{TopReX}\xspace}
\providecommand{\XDAQ} {{\textsc{xdaq}}\xspace}

%  Experiments
\newcommand {\DZERO}{D\O\xspace}     %etc.

% Measurements and units...

\newcommand{\de}{\ensuremath{^\circ}}
\newcommand{\ten}[1]{\ensuremath{\times \text{10}^\text{#1}}}
\newcommand{\unit}[1]{\ensuremath{\text{\,#1}}\xspace}
\newcommand{\mum}{\ensuremath{\,\mu\text{m}}\xspace}
\newcommand{\micron}{\ensuremath{\,\mu\text{m}}\xspace}
\newcommand{\cm}{\ensuremath{\,\text{cm}}\xspace}
\newcommand{\mm}{\ensuremath{\,\text{mm}}\xspace}
\newcommand{\mus}{\ensuremath{\,\mu\text{s}}\xspace}
\newcommand{\keV}{\ensuremath{\,\text{ke\hspace{-.08em}V}}\xspace}
\newcommand{\MeV}{\ensuremath{\,\text{Me\hspace{-.08em}V}}\xspace}
\newcommand{\GeV}{\ensuremath{\,\text{Ge\hspace{-.08em}V}}\xspace}
\newcommand{\TeV}{\ensuremath{\,\text{Te\hspace{-.08em}V}}\xspace}
\newcommand{\PeV}{\ensuremath{\,\text{Pe\hspace{-.08em}V}}\xspace}
\newcommand{\keVc}{\ensuremath{{\,\text{ke\hspace{-.08em}V\hspace{-0.16em}/\hspace{-0.08em}c}}}\xspace}
\newcommand{\MeVc}{\ensuremath{{\,\text{Me\hspace{-.08em}V\hspace{-0.16em}/\hspace{-0.08em}c}}}\xspace}
\newcommand{\GeVc}{\ensuremath{{\,\text{Ge\hspace{-.08em}V\hspace{-0.16em}/\hspace{-0.08em}c}}}\xspace}
\newcommand{\TeVc}{\ensuremath{{\,\text{Te\hspace{-.08em}V\hspace{-0.16em}/\hspace{-0.08em}c}}}\xspace}
\newcommand{\keVcc}{\ensuremath{{\,\text{ke\hspace{-.08em}V\hspace{-0.16em}/\hspace{-0.08em}c}^\text{2}}}\xspace}
\newcommand{\MeVcc}{\ensuremath{{\,\text{Me\hspace{-.08em}V\hspace{-0.16em}/\hspace{-0.08em}c}^\text{2}}}\xspace}
\newcommand{\GeVcc}{\ensuremath{{\,\text{Ge\hspace{-.08em}V\hspace{-0.16em}/\hspace{-0.08em}c}^\text{2}}}\xspace}
\newcommand{\TeVcc}{\ensuremath{{\,\text{Te\hspace{-.08em}V\hspace{-0.16em}/\hspace{-0.08em}c}^\text{2}}}\xspace}

\newcommand{\pbinv} {\mbox{\ensuremath{\,\text{pb}^\text{$-$1}}}\xspace}
\newcommand{\fbinv} {\mbox{\ensuremath{\,\text{fb}^\text{$-$1}}}\xspace}
\newcommand{\nbinv} {\mbox{\ensuremath{\,\text{nb}^\text{$-$1}}}\xspace}
\newcommand{\percms}{\ensuremath{\,\text{cm}^\text{$-$2}\,\text{s}^\text{$-$1}}\xspace}
\newcommand{\lumi}{\ensuremath{\mathcal{L}}\xspace}
\newcommand{\Lumi}{\ensuremath{\mathcal{L}}\xspace}%both upper and lower
%
% Need a convention here:
\newcommand{\LvLow}  {\ensuremath{\mathcal{L}=\text{10}^\text{32}\,\text{cm}^\text{$-$2}\,\text{s}^\text{$-$1}}\xspace}
\newcommand{\LLow}   {\ensuremath{\mathcal{L}=\text{10}^\text{33}\,\text{cm}^\text{$-$2}\,\text{s}^\text{$-$1}}\xspace}
\newcommand{\lowlumi}{\ensuremath{\mathcal{L}=\text{2}\times \text{10}^\text{33}\,\text{cm}^\text{$-$2}\,\text{s}^\text{$-$1}}\xspace}
\newcommand{\LMed}   {\ensuremath{\mathcal{L}=\text{2}\times \text{10}^\text{33}\,\text{cm}^\text{$-$2}\,\text{s}^\text{$-$1}}\xspace}
\newcommand{\LHigh}  {\ensuremath{\mathcal{L}=\text{10}^\text{34}\,\text{cm}^\text{$-$2}\,\text{s}^\text{$-$1}}\xspace}
\newcommand{\hilumi} {\ensuremath{\mathcal{L}=\text{10}^\text{34}\,\text{cm}^\text{$-$2}\,\text{s}^\text{$-$1}}\xspace}

% Some usual physics terms

\newcommand{\zp}{\ensuremath{\mathrm{Z}^\prime}\xspace}

% SM (still to be classified)

\newcommand{\kt}{\ensuremath{k_{\mathrm{T}}}\xspace}
\newcommand{\BC}{\ensuremath{{B_{\mathrm{c}}}}\xspace}
\newcommand{\bbarc}{\ensuremath{{\overline{b}c}}\xspace}
\newcommand{\bbbar}{\ensuremath{{b\overline{b}}}\xspace}
\newcommand{\ccbar}{\ensuremath{{c\overline{c}}}\xspace}
\newcommand{\JPsi}{\ensuremath{{J}/\psi}\xspace}
\newcommand{\bspsiphi}{\ensuremath{B_s \to \JPsi\, \phi}\xspace}
\newcommand{\AFB}{\ensuremath{A_\mathrm{FB}}\xspace}
\newcommand{\EE}{\ensuremath{e^+e^-}\xspace}
\newcommand{\MM}{\ensuremath{\mu^+\mu^-}\xspace}
\newcommand{\TT}{\ensuremath{\tau^+\tau^-}\xspace}
\newcommand{\wangle}{\ensuremath{\sin^{2}\theta_{\mathrm{eff}}^\mathrm{lept}(M^2_\mathrm{Z})}\xspace}
\newcommand{\ttbar}{\ensuremath{{t\overline{t}}}\xspace}
\newcommand{\stat}{\ensuremath{\,\text{(stat.)}}\xspace}
\newcommand{\syst}{\ensuremath{\,\text{(syst.)}}\xspace}
% these moved to similar defs
%\newcommand{\Etmiss}{\ensuremath{E_{\mathrm{T}\!{\rm miss}}}}
%\newcommand{\VEtmiss}{\ensuremath{{\vec E}_{\mathrm{T}\!{\rm miss}}}}

%%%  E-gamma definitions
\newcommand{\HGG}{\ensuremath{\mathrm{H}\to\gamma\gamma}}
\newcommand{\gev}{\GeV}
\newcommand{\GAMJET}{\ensuremath{\gamma + \mathrm{jet}}}
\newcommand{\PPTOJETS}{\ensuremath{\mathrm{pp}\to\mathrm{jets}}}
\newcommand{\PPTOGG}{\ensuremath{\mathrm{pp}\to\gamma\gamma}}
\newcommand{\PPTOGAMJET}{\ensuremath{\mathrm{pp}\to\gamma +
\mathrm{jet}
}}
\newcommand{\MH}{\ensuremath{\mathrm{M_{\mathrm{H}}}}}
\newcommand{\RNINE}{\ensuremath{\mathrm{R}_\mathrm{9}}}
\newcommand{\DR}{\ensuremath{\Delta\mathrm{R}}}

% Physics symbols ...

\newcommand{\PT}{\ensuremath{p_{\mathrm{T}}}\xspace}
\newcommand{\pt}{\ensuremath{p_{\mathrm{T}}}\xspace}
\newcommand{\ET}{\ensuremath{E_{\mathrm{T}}}\xspace}
\newcommand{\HT}{\ensuremath{H_{\mathrm{T}}}\xspace}
\newcommand{\et}{\ensuremath{E_{\mathrm{T}}}\xspace}
\newcommand{\Em}{\ensuremath{E\!\!\!/}\xspace}
\newcommand{\Pm}{\ensuremath{p\!\!\!/}\xspace}
\newcommand{\PTm}{\ensuremath{{p\!\!\!/}_{\mathrm{T}}}\xspace}
\newcommand{\ETm}{\ensuremath{E_{\mathrm{T}}^{\mathrm{miss}}}\xspace}
\newcommand{\MET}{\ensuremath{E_{\mathrm{T}}^{\mathrm{miss}}}\xspace}
\newcommand{\ETmiss}{\ensuremath{E_{\mathrm{T}}^{\mathrm{miss}}}\xspace}
\newcommand{\VEtmiss}{\ensuremath{{\vec E}_{\mathrm{T}}^{\mathrm{miss}}}\xspace}

%%%%%%
% From Albert
%

\newcommand{\ga}{\ensuremath{\gtrsim}}
\newcommand{\la}{\ensuremath{\lesssim}}
\newcommand{\swsq}{\ensuremath{\sin^2\theta_W}\xspace}
\newcommand{\cwsq}{\ensuremath{\cos^2\theta_W}\xspace}
\newcommand{\tanb}{\ensuremath{\tan\beta}\xspace}
\newcommand{\tanbsq}{\ensuremath{\tan^{2}\beta}\xspace}
\newcommand{\sidb}{\ensuremath{\sin 2\beta}\xspace}
\newcommand{\alpS}{\ensuremath{\alpha_S}\xspace}
\newcommand{\alpt}{\ensuremath{\tilde{\alpha}}\xspace}

\newcommand{\QL}{\ensuremath{Q_L}\xspace}
\newcommand{\sQ}{\ensuremath{\tilde{Q}}\xspace}
\newcommand{\sQL}{\ensuremath{\tilde{Q}_L}\xspace}
\newcommand{\ULC}{\ensuremath{U_L^C}\xspace}
\newcommand{\sUC}{\ensuremath{\tilde{U}^C}\xspace}
\newcommand{\sULC}{\ensuremath{\tilde{U}_L^C}\xspace}
\newcommand{\DLC}{\ensuremath{D_L^C}\xspace}
\newcommand{\sDC}{\ensuremath{\tilde{D}^C}\xspace}
\newcommand{\sDLC}{\ensuremath{\tilde{D}_L^C}\xspace}
\newcommand{\LL}{\ensuremath{L_L}\xspace}
\newcommand{\sL}{\ensuremath{\tilde{L}}\xspace}
\newcommand{\sLL}{\ensuremath{\tilde{L}_L}\xspace}
\newcommand{\ELC}{\ensuremath{E_L^C}\xspace}
\newcommand{\sEC}{\ensuremath{\tilde{E}^C}\xspace}
\newcommand{\sELC}{\ensuremath{\tilde{E}_L^C}\xspace}
\newcommand{\sEL}{\ensuremath{\tilde{E}_L}\xspace}
\newcommand{\sER}{\ensuremath{\tilde{E}_R}\xspace}
\newcommand{\sFer}{\ensuremath{\tilde{f}}\xspace}
\newcommand{\sQua}{\ensuremath{\tilde{q}}\xspace}
\newcommand{\sUp}{\ensuremath{\tilde{u}}\xspace}
\newcommand{\suL}{\ensuremath{\tilde{u}_L}\xspace}
\newcommand{\suR}{\ensuremath{\tilde{u}_R}\xspace}
\newcommand{\sDw}{\ensuremath{\tilde{d}}\xspace}
\newcommand{\sdL}{\ensuremath{\tilde{d}_L}\xspace}
\newcommand{\sdR}{\ensuremath{\tilde{d}_R}\xspace}
\newcommand{\sTop}{\ensuremath{\tilde{t}}\xspace}
\newcommand{\stL}{\ensuremath{\tilde{t}_L}\xspace}
\newcommand{\stR}{\ensuremath{\tilde{t}_R}\xspace}
\newcommand{\stone}{\ensuremath{\tilde{t}_1}\xspace}
\newcommand{\sttwo}{\ensuremath{\tilde{t}_2}\xspace}
\newcommand{\sBot}{\ensuremath{\tilde{b}}\xspace}
\newcommand{\sbL}{\ensuremath{\tilde{b}_L}\xspace}
\newcommand{\sbR}{\ensuremath{\tilde{b}_R}\xspace}
\newcommand{\sbone}{\ensuremath{\tilde{b}_1}\xspace}
\newcommand{\sbtwo}{\ensuremath{\tilde{b}_2}\xspace}
\newcommand{\sLep}{\ensuremath{\tilde{l}}\xspace}
\newcommand{\sLepC}{\ensuremath{\tilde{l}^C}\xspace}
\newcommand{\sEl}{\ensuremath{\tilde{e}}\xspace}
\newcommand{\sElC}{\ensuremath{\tilde{e}^C}\xspace}
\newcommand{\seL}{\ensuremath{\tilde{e}_L}\xspace}
\newcommand{\seR}{\ensuremath{\tilde{e}_R}\xspace}
\newcommand{\snL}{\ensuremath{\tilde{\nu}_L}\xspace}
\newcommand{\sMu}{\ensuremath{\tilde{\mu}}\xspace}
\newcommand{\sNu}{\ensuremath{\tilde{\nu}}\xspace}
\newcommand{\sTau}{\ensuremath{\tilde{\tau}}\xspace}
\newcommand{\Glu}{\ensuremath{g}\xspace}
\newcommand{\sGlu}{\ensuremath{\tilde{g}}\xspace}
\newcommand{\Wpm}{\ensuremath{W^{\pm}}\xspace}
\newcommand{\sWpm}{\ensuremath{\tilde{W}^{\pm}}\xspace}
\newcommand{\Wz}{\ensuremath{W^{0}}\xspace}
\newcommand{\sWz}{\ensuremath{\tilde{W}^{0}}\xspace}
\newcommand{\sWino}{\ensuremath{\tilde{W}}\xspace}
\newcommand{\Bz}{\ensuremath{B^{0}}\xspace}
\newcommand{\sBz}{\ensuremath{\tilde{B}^{0}}\xspace}
\newcommand{\sBino}{\ensuremath{\tilde{B}}\xspace}
\newcommand{\Zz}{\ensuremath{Z^{0}}\xspace}
\newcommand{\sZino}{\ensuremath{\tilde{Z}^{0}}\xspace}
\newcommand{\sGam}{\ensuremath{\tilde{\gamma}}\xspace}
\newcommand{\chiz}{\ensuremath{\tilde{\chi}^{0}}\xspace}
\newcommand{\chip}{\ensuremath{\tilde{\chi}^{+}}\xspace}
\newcommand{\chim}{\ensuremath{\tilde{\chi}^{-}}\xspace}
\newcommand{\chipm}{\ensuremath{\tilde{\chi}^{\pm}}\xspace}
\newcommand{\Hone}{\ensuremath{H_{d}}\xspace}
\newcommand{\sHone}{\ensuremath{\tilde{H}_{d}}\xspace}
\newcommand{\Htwo}{\ensuremath{H_{u}}\xspace}
\newcommand{\sHtwo}{\ensuremath{\tilde{H}_{u}}\xspace}
\newcommand{\sHig}{\ensuremath{\tilde{H}}\xspace}
\newcommand{\sHa}{\ensuremath{\tilde{H}_{a}}\xspace}
\newcommand{\sHb}{\ensuremath{\tilde{H}_{b}}\xspace}
\newcommand{\sHpm}{\ensuremath{\tilde{H}^{\pm}}\xspace}
\newcommand{\hz}{\ensuremath{h^{0}}\xspace}
\newcommand{\Hz}{\ensuremath{H^{0}}\xspace}
\newcommand{\Az}{\ensuremath{A^{0}}\xspace}
\newcommand{\Hpm}{\ensuremath{H^{\pm}}\xspace}
\newcommand{\sGra}{\ensuremath{\tilde{G}}\xspace}
\newcommand{\mtil}{\ensuremath{\tilde{m}}\xspace}
\newcommand{\rpv}{\ensuremath{\rlap{\kern.2em/}R}\xspace}
\newcommand{\LLE}{\ensuremath{LL\bar{E}}\xspace}
\newcommand{\LQD}{\ensuremath{LQ\bar{D}}\xspace}
\newcommand{\UDD}{\ensuremath{\overline{UDD}}\xspace}
\newcommand{\Lam}{\ensuremath{\lambda}\xspace}
\newcommand{\Lamp}{\ensuremath{\lambda'}\xspace}
\newcommand{\Lampp}{\ensuremath{\lambda''}\xspace}
\newcommand{\spinbd}[2]{\ensuremath{\bar{#1}_{\dot{#2}}}\xspace}

\newcommand{\MD}{\ensuremath{{M_\mathrm{D}}}\xspace}% ED mass
\newcommand{\Mpl}{\ensuremath{{M_\mathrm{Pl}}}\xspace}% Planck mass
\newcommand{\Rinv} {\ensuremath{{R}^{-1}}\xspace}

%%%%%%%%%%%%%%%%%%%%%%%%%%%%%%%%%%%%%%%%%%%%%%%%%%%%%%%%%%%%%%%%%%%%
%
% Hyphenations (only need to add here if you get a nasty word break)
%
\hyphenation{en-viron-men-tal}%    just an example

%%%%%%%%%%%%%%%  Title page %%%%%%%%%%%%%%%%%%%%%%%%
\cmsNoteHeader{09-016}
\title{Alignment of the CMS Muon System \\ with Cosmic-Ray and Beam-Halo Muons}% Force line breaks with \\

%Author is always "The CMS Collaboration" for PAS, so author, etc will be ignored

% please supply the date in yyyy/mm/dd format. Today has been
% redefined to do so, but it should be fixed as of the final release date.
\date{\today}

% note that you cannot use \verb in the abstract text
\abstract{The CMS muon system has been aligned using cosmic-ray muons
collected in 2008 and beam-halo muons from the 2008 LHC circulating
beam tests.  After alignment, the resolution of the most sensitive
coordinate is 80~microns for the relative positions of superlayers in
the same barrel chamber and 270~microns for the relative positions of
endcap chambers in the same ring structure.  The resolution on the
position of the central barrel chambers relative to the tracker is
comprised between two extreme estimates, 200 and 700~microns,
provided by two complementary studies.  With minor modifications, the
alignment procedures can be applied using muons from LHC collisions,
leading to additional significant improvements.}

% these need to be filled in by hand and should (MUST) match the info
% in the TeX equivalents less the TeX markup
\hypersetup{%
pdfauthor={CMS Collaboration},%
pdftitle={Alignment of the CMS Muon System with Cosmic-Ray and Beam-Halo Muons},%
pdfsubject={CMS},%
pdfkeywords={CMS, detectors, CRAFT, muons, alignment, tracks}}

\maketitle %maketitle comes after all the front information has been supplied

%%%%%%%%%%%%%%%%%%%%%%%%%%%%%%%%  Begin text %%%%%%%%%%%%%%%%%%%%%%%%%%%%%

%---------------------------------------------------------------------
\section{Introduction}

The primary goal of the Compact Muon Solenoid (CMS)
experiment~\cite{:2008zzk} is to explore particle physics at the TeV
energy scale exploiting the proton-proton collisions delivered by the
Large Hadron Collider (LHC)~\cite{Evans:2008zzb}.  The CMS experiment
features a large muon tracking system for identifying muons and
reconstructing their momenta.  As with all tracking systems, the
momentum resolution of reconstructed tracks depends in part on the
alignment of detector components in space: three translational and
three rotational degrees of freedom for each component.  The alignment
procedure provides corrections to the relationships between local
detector-bound coordinate frames and a single, global frame for all
CMS tracking systems.

\begin{figure}[p]
\includegraphics[width=\linewidth]{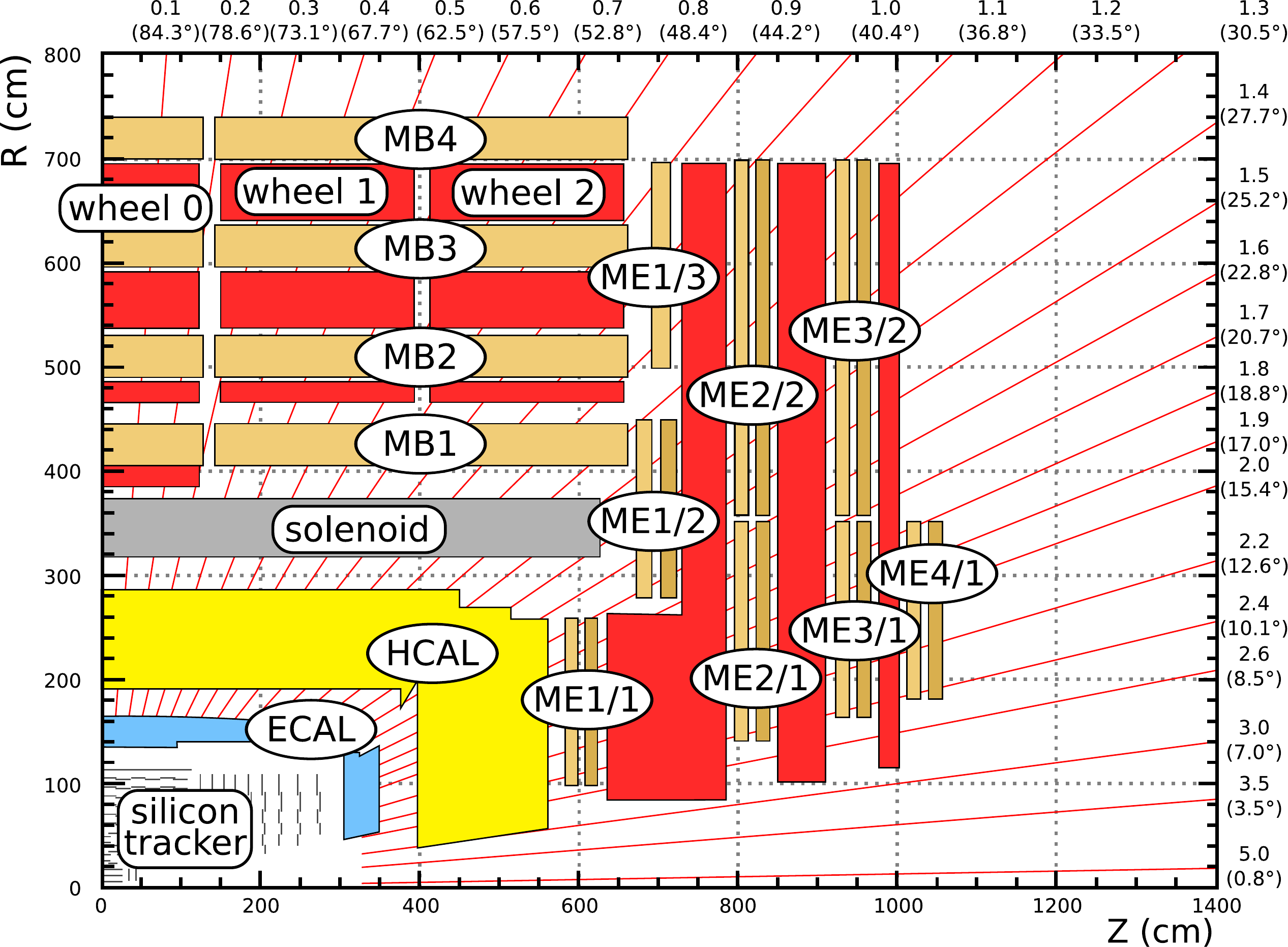}

\caption{Quarter-view of CMS with labeled muon barrel (MB) and endcap
(ME) stations.  The steel yoke is represented by darkly shaded (red) blocks
between the muon chambers.  Pseudorapidities and polar angles are
indicated on the top and right edges of the diagram. \label{fig:muon_system_labeled}}
\end{figure}

\begin{figure}[p]
\includegraphics[width=\linewidth]{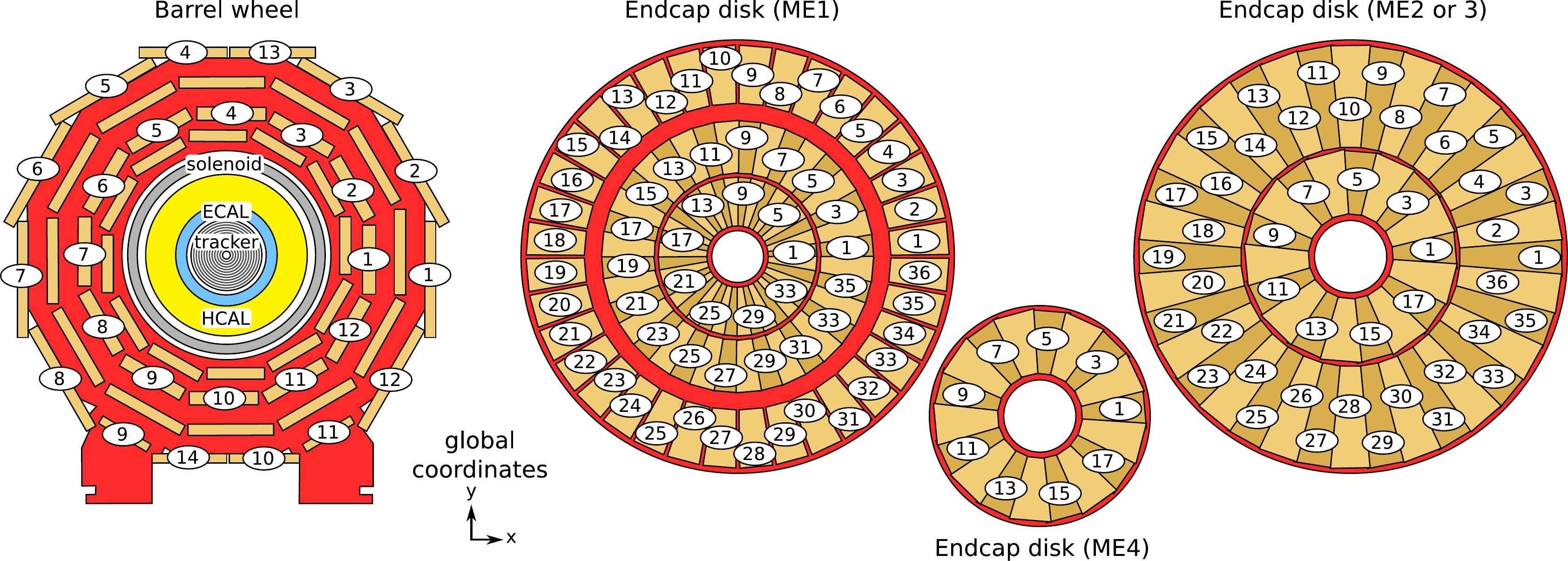}

\caption{Transverse slices of CMS: the central barrel wheel and
representative endcap disks, indicating the numbering of the DT azimuthal sectors in
the barrel and the CSC chamber numbers in the endcaps. \label{fig:muoneverything}}
\end{figure}

The muon system consists of hundreds of independent tracking chambers
mounted within the CMS magnetic field return yoke.  Three technologies
are employed: Drift Tube (DT) chambers on the five modular wheels of
the barrel section, Cathode
Strip Chambers (CSC) on the six endcap disks (illustrated in
Figs.~\ref{fig:muon_system_labeled} and \ref{fig:muoneverything}) and
Resistive Plate Chambers (RPC) throughout.  The DTs and CSCs are
sufficiently precise to contribute to the momentum resolution of
high-momentum muons (several hundred GeV/$c$) assuming that these
chambers are well-aligned relative to the CMS tracker, a one-meter
radius silicon strip and pixel detector.  Between the tracker and the
muon system are electromagnetic and hadronic calorimeters (ECAL and
HCAL, respectively) for particle identification and energy
measurement, as well as the solenoid coil for producing an operating
magnetic field strength of~3.8~T in which to measure charged-particle
momenta (all shown in Fig.~\ref{fig:muon_system_labeled}).

The CMS collaboration is developing multiple techniques to align the
DT and CSC chambers and their internal layers.  Photogrammetry and
in-situ measurement devices~\cite{ref:hardware_alignment} provide
real-time monitoring of potential chamber movements on short
timescales and measurements of degrees of freedom to which tracks are
only weakly sensitive.  Track-based alignment, the subject of this
paper, optimizes component positions for a given set of tracks,
directly relating the active elements of the detectors traversed by
the charged particles in a shared coordinate frame.  Methods using
tracks are employed both to align nearby components relative to one
another and to align all muon chambers relative to the tracker.

A challenge to track-based alignment in the CMS muon system is the
presence of large quantities of material between the chambers.  As a
central design feature of the detector, 20--60~cm layers of steel
are sandwiched between the chambers to concentrate the magnetic field
and absorb beam-produced hadrons.  Consequently, uncertainties in
track trajectories become significant as muons propagate through the
material, making it necessary to develop alignment procedures that are
insensitive to scattering, even though typical deviations in the muon
trajectories (3--8~mm) are large compared to the intrinsic spatial
resolution (100--300~$\mu$m).  Two types of approaches are presented
in this paper: the relative alignment of nearby structures, which avoids
extrapolation of tracks through material but does not relate distant
coordinate frames to each other, and the alignment using tracks
reconstructed in the tracker, which allows for a more sophisticated
treatment of propagation effects by simplifying the interdependence of
alignment parameters.

This paper begins with a brief overview of the geometry of the muon
system and conventions to be used thereafter
(Section~\ref{sec:geometry}), followed by presentations of three
alignment procedures:
{\renewcommand{\labelenumi}{(\alph{enumi})}
\begin{enumerate}
\item internal alignment of layers within DT chambers using a
combination of locally fitted track segments and survey measurements
(Section~\ref{sec:localdt});
\item alignment of groups of overlapping CSC chambers relative to one
another, using only locally fitted track segments
(Section~\ref{sec:localcsc});
\item alignment of each chamber relative to the tracker, using
the tracks from the tracker, propagated to the muon system with a
detailed map of the magnetic field and material distribution of CMS
(Section~\ref{sec:global_muon_alignment}).
\end{enumerate}}

Procedure (c), above, completes the alignment, relating all
local coordinate frames to a shared frame.  Its performance is greatly
improved by supplying internally aligned chambers from procedure (a),
such that only rigid-body transformations of whole chambers need to be
considered.  Procedures (b) and (c) both align CSC chambers relative
to one another, but in different ways: (b) does not need many tracks,
only about 1000 per chamber, to achieve high precision, and
(c) additionally links the chambers to the tracker.

With the first LHC collisions, groups of CSCs will be interaligned
using (b) and these rigid-body groups will be aligned relative to the
tracker with (c).  As more data become available, comparisons of
results from (b) and (c) yield highly sensitive tests of systematic
errors in (c).

Although the ideal tracks for these procedures are muons from LHC
collisions, this paper focuses on application of the procedures using
currently available data, namely cosmic rays (a and c) and beam-halo
muons from circulating LHC beam tests in September 2008 (b).  In
particular, (c) requires a magnetic field to select high-quality,
high-momentum muons and concurrent operation of the tracker and muon
systems.  The CMS Collaboration conducted a month-long data-taking
exercise known as the Cosmic Run At Four Tesla (CRAFT) during
October--November 2008, with the goal of commissioning the experiment
for extended operation~\cite{ref:CRAFTGeneral}.  With all installed
detector systems participating, CMS recorded 270~million cosmic-ray
triggered events with the solenoid at its nominal axial field strength
of 3.8~T.  Due to geometric limitations imposed by the primarily
vertical distribution of cosmic rays underground, (c) is performed with
only a subset of DT chambers using CRAFT data, though the procedure
will apply similarly to CSC chambers, once a large sample of
inclined tracks becomes available.

The formalism and results of each procedure are presented together.
Details of the data transfer and the computing model which were used
to implement these procedures are described in
Ref.~\cite{ref:workflow}.

\section{Geometry of the Muon System and Definitions}
\label{sec:geometry}

Muon chambers are independent, modular track detectors, each
containing 6--12 measurement layers, sufficient to determine the
position and direction of a passing muon from the intersections of its
trajectory with the layer planes (``hits'').  The DT layers are
oriented nearly perpendicular to lines radially projected from the
beamline, and CSC layers are perpendicular to lines parallel with the
beamline.  Hits are initially expressed in a local coordinate frame
$(x, y, z)$ defined by the layers: $z=0$ is the plane of the layer and
$x$ is the more precisely measured (or the only measured) of the two
plane coordinates.  On CSC layers, the most precise measurement is
given by cathode strips, which fan radially from the
beamline~\cite{ref:csc_resolution}.  Defining ``local $r\phi$'' as the
curvilinear coordinate orthogonal to the strips at all points, $x$ and
local $r\phi$ coincide only at the center of each CSC layer.

A semi-local coordinate system for the entire chamber is defined with $x$,
$y$, and $z$ axes nominally parallel to the layers' axes, but with a
single origin.  Within this common frame, the positions of
hits from different layers can be related to each other and combined
by a linear fit into segments with position $(\bar{x}, \bar{y})$
and direction $(\frac{\textrm{d}x}{\textrm{d}z}, \frac{\textrm{d}y}{\textrm{d}z})$.  The nominal $x$
direction of every chamber is perpendicular to the beamline and
radial projections from the beamline.

\begin{figure}
\centering
\subfigure[]{\includegraphics[height=6 cm]{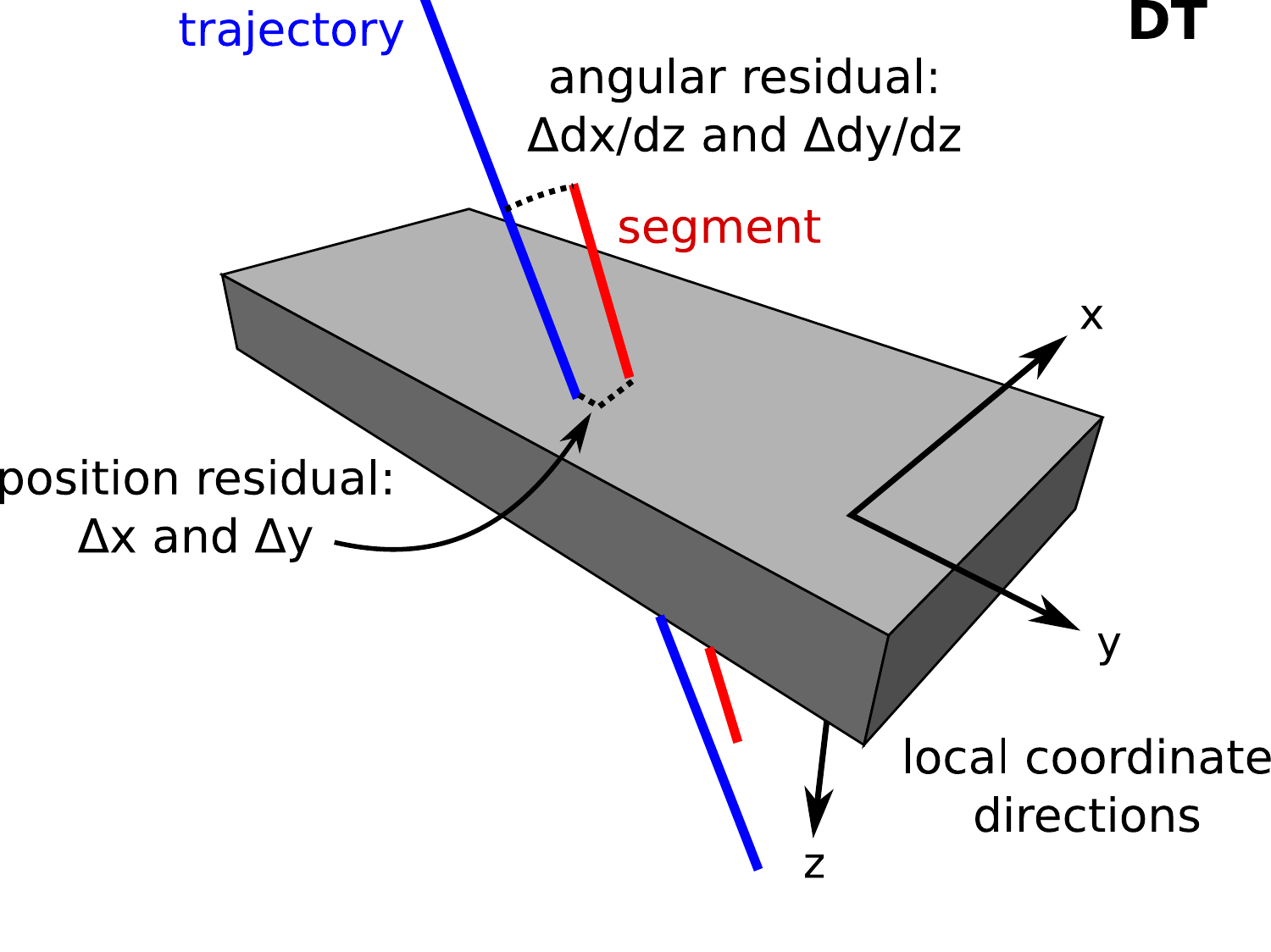} \label{fig:coordinates-a}}
\hfill \subfigure[]{\includegraphics[height=6 cm]{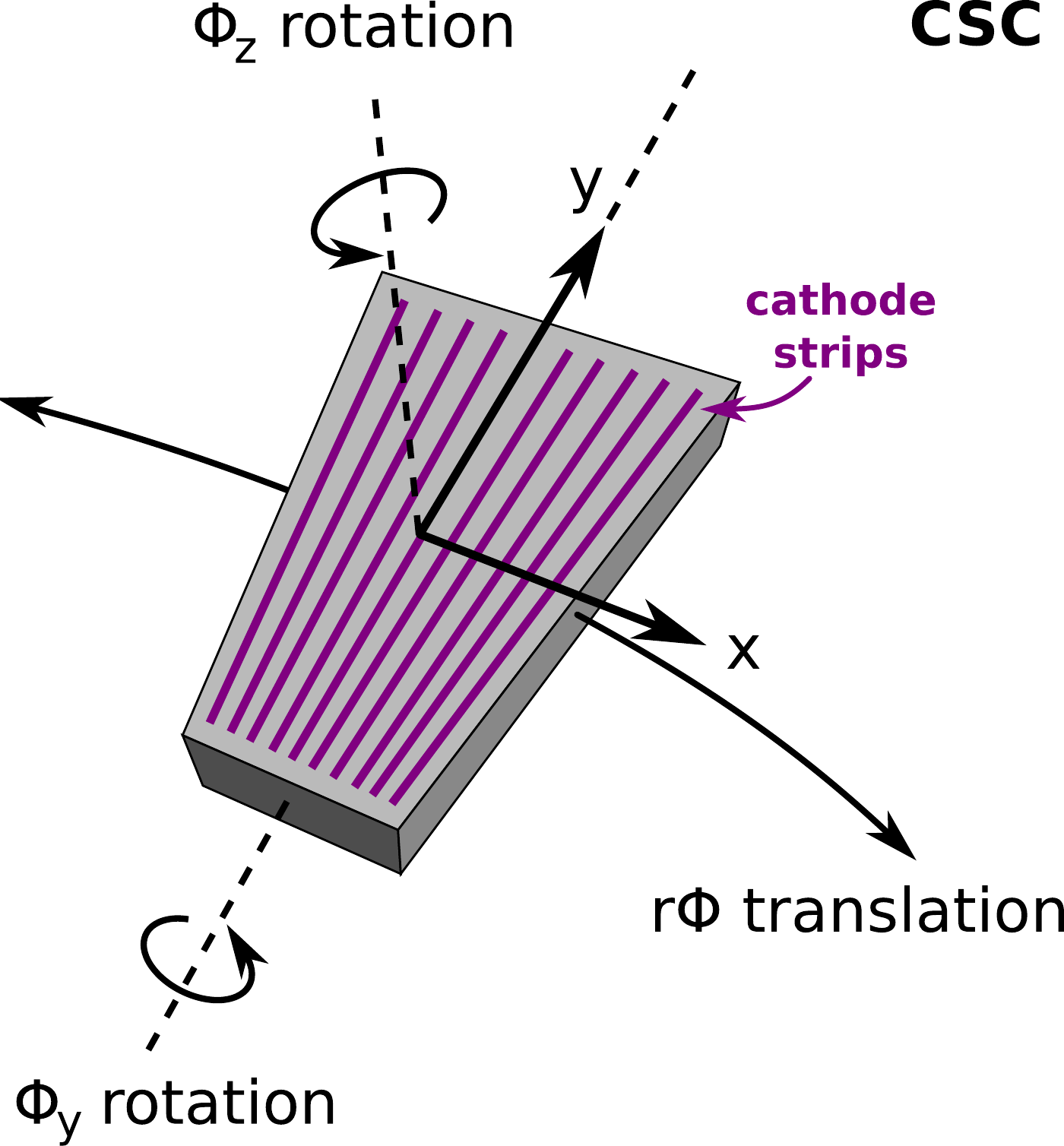} \label{fig:coordinates-b}}

\caption{(a) Coordinates and residuals for a DT chamber. (b) Coordinates and alignment parameters for a CSC chamber.}
\end{figure}

Residuals are differences between the predicted particle
trajectories and the muon chamber data.  Residuals can have as few as
one component, from layers that measure only one dimension in the
measurement plane, and as many as four components, $\Delta x$, $\Delta y$,
$\Delta \frac{\textrm{d}x}{\textrm{d}z}$, $\Delta \frac{\textrm{d}y}{\textrm{d}z}$, from linear fits to
all residuals in a chamber, as illustrated in
Fig.~\ref{fig:coordinates-a}.

To compare chamber positions to each other and to the position of the
tracker, consider global coordinates $(X, Y, Z)$ with the
origin at the symmetry center of CMS, $Z$ axis directed anti-clockwise along the
beamline, $Y$ upward, and $X$ horizontally toward the center
of the LHC ring.  Local coordinate systems are related to one another
through an Alignment Integration Frame (AIF) that only approximates these
global coordinates.  It is not necessary to tightly control the
definition of the AIF, as global translations
and rotations of the whole system do not affect any physics
measurements.  Specific coordinate frames, at intermediate levels
between layers and chambers and between chambers and global, are
introduced in this paper as needed.

Corrections to positions and orientations of the layers and chambers
are denoted as $\delta_x$, $\delta_y$, $\delta_z$, $\delta_{\phi_x}$,
$\delta_{\phi_y}$, and $\delta_{\phi_z}$, with $x$, $y$, and $z$ expressed in
local coordinates of the layer or chamber, and $\phi_x$, $\phi_y$,
and $\phi_z$ as rotations around the corresponding axis; alignment
corrections are small enough, $\mathcal{O}(\mbox{mrad})$, to
approximately commute.  For CSC chambers, $\Delta(r\phi)$,
$\Delta \frac{\textrm{d}(r\phi)}{\textrm{d}z}$, and $\delta_{r\phi}$
are more appropriate than $\Delta x$,
$\Delta \frac{\textrm{d}x}{\textrm{d}z}$, and $\delta_x$,
respectively, to take advantage of the precision of the cathode
strips.  The most sensitive CSC alignment parameters are illustrated
in Fig.~\ref{fig:coordinates-b}.

DT layers are grouped into four-layer ``superlayers,'' each of which
measures one coordinate.  Most DT chambers contain three~superlayers, and
the middle one, superlayer~2, is oriented a 90$^\circ$ angle with
respect to superlayers~1 and 3 to measure $y$ positions in the chamber
coordinate frame.  DT chambers farthest from the interaction point
(MB4 in Fig.~\ref{fig:muon_system_labeled}) contain only superlayers~1
and 3, and are therefore insensitive to $y$ and $\frac{\textrm{d}y}{\textrm{d}z}$.  CSC
chambers contain six identical layers.

Chambers are grouped into ``stations'' by their distance from the
interaction region, named MB1--MB4 in the barrel and ME$\pm$1/1,
$\pm$1/2, $\pm$1/3, $\pm$2/1, $\pm$2/2, $\pm$3/1, $\pm$3/2, and
$\pm$4/1 in the endcaps, labeled in
Fig.~\ref{fig:muon_system_labeled}.  Azimuthal positions are called
``sectors'' in the barrel and simply ``chamber number'' in the
endcaps, and these are labeled in Fig.~\ref{fig:muoneverything}.

\section{Internal Alignment of DT Chambers}
\label{sec:localdt}

The layers and superlayers of DT chambers are aligned and analyzed in
two steps.  First, all layers in the chamber are aligned using a general Millepede
algorithm~\cite{Blobel:2006yh}, simultaneously optimizing alignment
parameters and segment parameters, subject to constraints from a
survey performed during construction.  In the second step, the
alignment of superlayers 1 and 3 are checked by fitting
segments in each superlayer separately and measuring alignment
errors from deviations between pairs of superlayer segments.

\subsection{Layer Alignment}
\label{sec:localdt1}

Measured trajectories within a single chamber do not suffer
from uncertainties due to scattering in the steel between chambers.  Segments,
determined from linear fits to hits in one chamber only, are therefore used to
determine the alignment of layers inside the chamber.  The segment parameters
depend on the layer alignment parameters within the chamber, so
both must be resolved in a combined fit.

Without external constraints, the combined fit does not have a unique
optimum, as it is insensitive to global distortions of the chamber.
Consider a shear of the chamber that translates each layer in $x$ by
an amount proportional to its $z$ position and an equal shear of all
segment angles: this new geometry has exactly the same residuals as the
unsheared geometry.

It is therefore necessary to add external survey measurements to
constrain the fit; these are taken from two sources: (1)~measurements
of the wire end-pins taken during superlayer
construction, and (2)~photographs of reflective targets
on the exterior of the superlayer (``photogrammetry''), taken during
chamber construction.  The averages of wire positions for all wires in
each layer provide measurements of the $x$ positions of the layers:
RMS deviations from design geometry are 100~$\mu$m with 30--40~$\mu$m
uncertainties.  RMS deviations for the photogrammetry measurements
are 200~$\mu$m in $x$ and $y$, 500~$\mu$m in $z$, and 150~$\mu$rad in
$\phi_x$, $\phi_y$, and~$\phi_z$.

The alignment of a tracking system subject to survey constraints can
be expressed as the minimum of an objective function with terms
derived from both tracks and survey.  The objective function is
\begin{multline}
\chi^2 = \sum_i^{\mbox{\scriptsize layers}} \sum_j^{\mbox{\scriptsize tracks}}
\left(\Delta \vec{x}_{ij} - A_j \cdot \vec{\delta}_i - B_i \cdot \delta \vec{p}_j\right)^T
({\sigma_{\mbox{\scriptsize hit}}}^2)_{ij}^{-1} \left(\Delta \vec{x}_{ij} - A_j \cdot \vec{\delta}_i - B_i \cdot \delta \vec{p}_j\right) \\
+ \sum_i^{\mbox{\scriptsize layers}} \sum_k^{\mbox{\scriptsize targets}}
\left(\Delta \vec{\xi}_k - C_{ik} \cdot \vec{\delta}_i\right)^T
({\sigma_{\mbox{\scriptsize survey}}}^2)_k^{-1}
\left(\Delta \vec{\xi}_k - C_{ik} \cdot \vec{\delta}_i\right)
+ \lambda \left|\sum_i^{\mbox{\scriptsize layers}} \vec{\delta}_i\right|^2 \quad\mbox{,}
\label{eqn:chi2millepede}
\end{multline}
where
\begin{itemize}
\item $\Delta \vec{x}_{ij}$ is the residual on layer $i$ from track $j$ (one-dimensional and in layer coordinates for the layer alignment case);
\item $\vec{\delta}_i =
(\delta_x, \delta_y, \delta_z, \delta_{\phi_x}, \delta_{\phi_y}, \delta_{\phi_z})$ is a vector of alignment corrections for layer $i$;
\item $A_j = \left(\begin{array}{c c c c c c} 1 & 0 & -\frac{\textrm{d}x}{\textrm{d}z}_j & -y_j \frac{\textrm{d}x}{\textrm{d}z}_j & x_j \frac{\textrm{d}x}{\textrm{d}z}_j & -y_j \end{array}\right)$ is a 1$\times$6 matrix transforming alignment parameter errors to residuals, dependent on the layer intersection $(x_j, y_j)$ and entrance angle $\frac{\textrm{d}x}{\textrm{d}z}_j$ of track $j$;
\item $\delta \vec{p}_j$ is a vector of corrections to the segment parameters;
\item $B_i$ is a matrix transforming variation of segment parameters
into residuals on layer $i$ (including a projection to the measured direction in the layer coordinate system);
\item $({\sigma_{\mbox{\scriptsize hit}}}^2)_{ij}^{-1}$ is the inverse
of the covariance matrix of the uncertainty in the hit on layer $i$,
track $j$ (single-valued in the layer alignment case);
\item $\Delta \vec{\xi}_k$ is the difference between the nominal and
measured position of survey target $k$;
\item $C_{ik}$ is a matrix transforming alignment errors in layer $i$ to corrections in the position of survey target $k$;
\item $({\sigma_{\mbox{\scriptsize survey}}}^2)_k^{-1}$ is the inverse
of the covariance matrix of measurement $k$;
\item $\lambda \left| \sum_i \vec{\delta}_i \right|^2$ is a Lagrange multiplier to inhibit translations
and rotations of the chamber.
\end{itemize}
Segments are modeled as straight lines because the magnetic
field is negligible inside the DT chambers.  Survey targets from wire
measurements are one-dimensional (the other two coordinates are given
zero inverse uncertainties in $({\sigma_{\mbox{\scriptsize
survey}}}^2)_k^{-1}$) and photogrammetry targets are included as
individual survey constraints.  Wire measurements apply only to
relative positions of layers within their parent superlayer, and
photogrammetry constraints apply to all layers in a superlayer as a
group.  This is expressed in $C_{ik} \cdot \vec{\delta}_i$ with terms
such as $\vec{\delta}_i - [\vec{\delta}_1 + \vec{\delta}_2
+ \vec{\delta}_3 + \vec{\delta}_4]/4$ to constrain layer $i$ in a
superlayer consisting of layers 1--4.  Errors in survey measurements
are assumed to be uncorrelated.

The $\chi^2$ is quadratic in its parameters, so it is solved by the matrix
inversion method.  All layers were aligned in $\delta_x$, $\delta_{\phi_x}$,
$\delta_{\phi_y}$, and $\delta_{\phi_z}$, using about $20\,000$ cosmic
ray tracks per chamber.  The RMS of the corrections is 116~$\mu$m,
58~$\mu$rad, 63~$\mu$rad, and 49~$\mu$rad, respectively, and will be
studied in more detail in the next section.

\subsection{Test of Superlayer Alignment}
\label{sec:superlayertest}

To cross-check the alignment results in superlayers~1 and 3, the
track-based data are compared with an independent set of photogrammetry
measurements.  The four hits in each superlayer define a superlayer
segment with a one-dimensional position and slope, and segments from
different superlayers must match at a common plane.  Segment residuals,
$\Delta x_{\mbox{\scriptsize S}}$, are the mis-match of these superlayer
segments, and are sensitive to relative positions of the superlayers as well as their
internal layer alignment parameters.

Segment residuals can be used to align superlayers~1 and 3 relative to
one another without needing external constraints.  The first term in
Eq.~(\ref{eqn:chi2millepede}) becomes
\begin{equation}
{\chi_{\mbox{\scriptsize S}}}^2 =
\sum_j^{\mbox{\scriptsize tracks}}
\left[\Delta {x_{\mbox{\scriptsize S}}}_j
- \bigg(\begin{array}{c c c}
1 & -\dfrac{\textrm{d}x}{\textrm{d}z}_j & x_j \dfrac{\textrm{d}x}{\textrm{d}z}_j \\
\end{array}\bigg) \cdot \left(\begin{array}{c}
\delta_{x_{\mbox{\scriptsize S}}} \\ \delta_{z_{\mbox{\scriptsize S}}} \\ \delta_{{\phi_y}_{\mbox{\scriptsize S}}} \\
\end{array} \right) \right]^2 \, \frac{1}{(\sigma_{{\Delta x_{\mbox{\scriptsize S}}}})_j^2} \quad\mbox{,}
\label{eqn:superlayerAlignment}
\end{equation}
with no explicit dependence on track parameter corrections
$\delta \vec{p}_j$.  The minimization of
Eq.~(\ref{eqn:superlayerAlignment}) has a unique solution in
$\delta_{x_{\mbox{\scriptsize S}}}$, $\delta_{z_{\mbox{\scriptsize
S}}}$, and $\delta_{{\phi_y}_{\mbox{\scriptsize S}}}$.  It is not
susceptible to shear, for example, because the slopes of the
segments are fixed by the layers in each superlayer.  Access to
additional alignment parameters would require knowledge of $y$ and
$\frac{\textrm{d}y}{\textrm{d}z}$ which are not available in superlayers~1 and 3.  The
position of superlayer~2 cannot be determined in this way because it
is the only $y$-measuring detector in a DT chamber.

\begin{figure}
\centering
\subfigure{\includegraphics[width=0.45\linewidth]{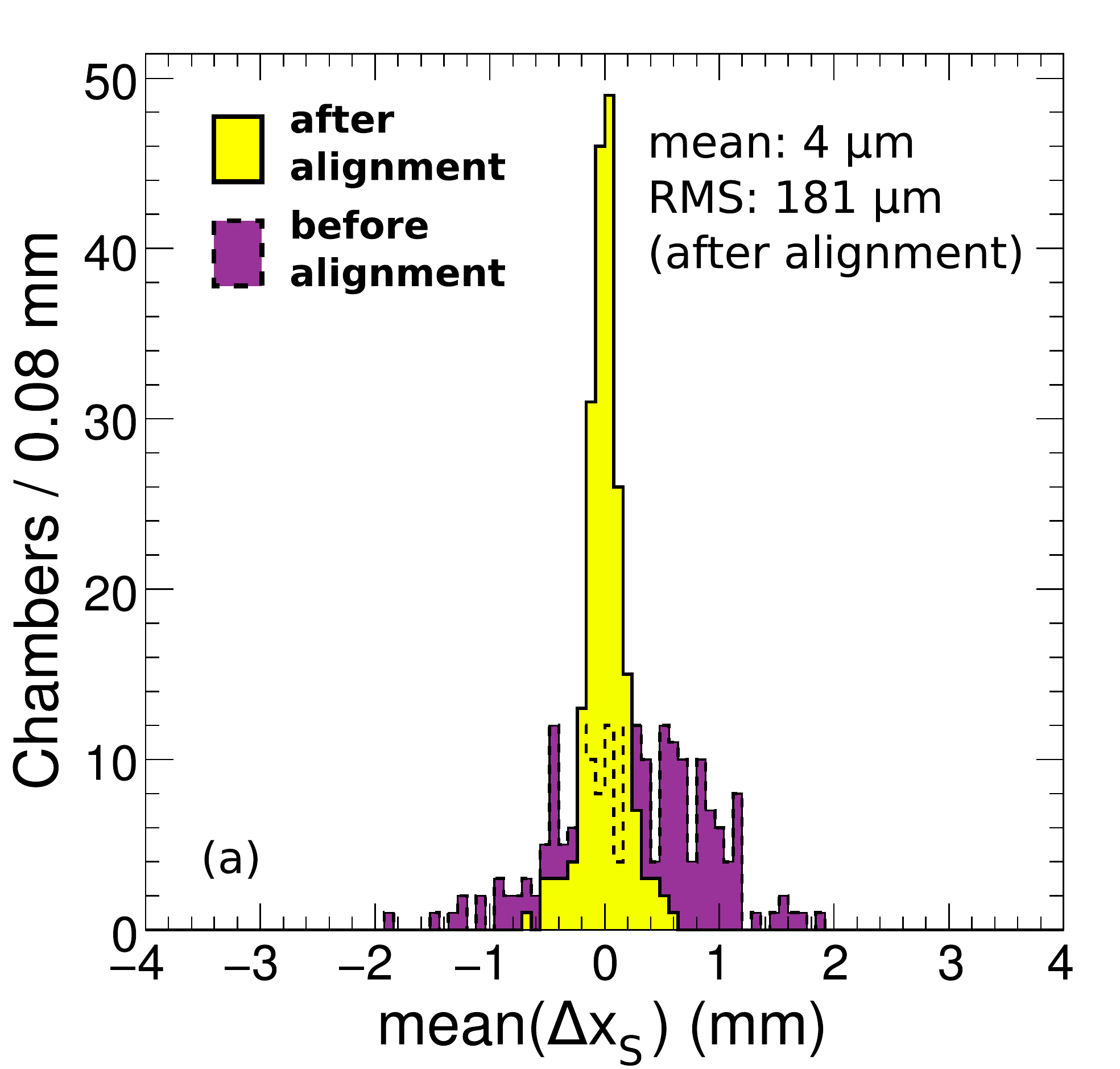}\label{fig:residualsStandaloneAlignment}}
\subfigure{\includegraphics[width=0.45\linewidth]{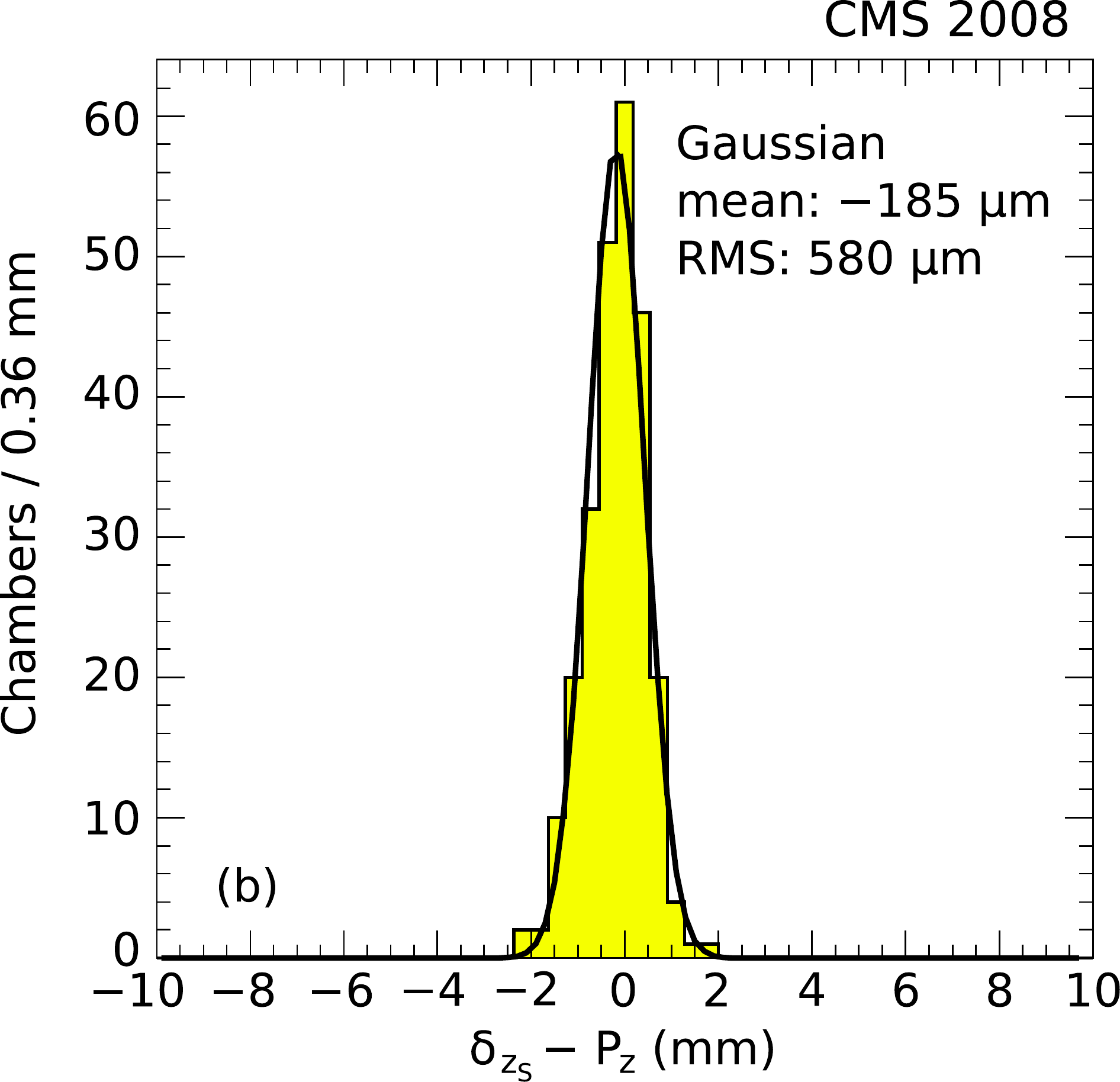}\label{fig:surveyvstracks}}
\caption{(a) Distribution of means of DT segment residuals for all
chambers, before alignment presented in a dark/dashed histogram and
after alignment in a light histogram.  (b)~Difference in $z$ of
superlayers as measured by tracks ($\delta_{z_{\mbox{\scriptsize
S}}}$) and photogrammetry ($P_z$).}
\end{figure}

To check the partially survey-based layer alignment from
Section~\ref{sec:localdt1}, segment residuals before and after
layer alignment are plotted in Fig.~\ref{fig:residualsStandaloneAlignment}.  Each
entry in the histogram is the mean of the $\Delta x_{\mbox{\scriptsize
S}}$ distribution for a single chamber.  After layer alignment,
segment positions and angles are more consistent between superlayers~1
and 3, leading to better matching at the common plane.

To verify the photogrammetry superlayer positions, 
$\delta_{x_{\mbox{\scriptsize S}}}$, $\delta_{z_{\mbox{\scriptsize
S}}}$, and $\delta_{{\phi_y}_{\mbox{\scriptsize S}}}$ from a
minimization of Eq.~(\ref{eqn:superlayerAlignment}) are directly compared with the
photogrammetry measurements $P_x$, $P_z$, and $P_{\phi_y}$ of the
relative superlayer positions.
Figure~\ref{fig:surveyvstracks} presents differences between the
$\delta_{z_{\mbox{\scriptsize S}}}$ and $P_z$ for each chamber.
Typical values of $\delta_{z_{\mbox{\scriptsize S}}}$ are 1--1.5~mm
due to a glue layer not included in the design geometry, and they
agree with $P_z$ on a per-chamber basis with 580~$\mu$m accuracy.
Agreement of $\delta_{x_{\mbox{\scriptsize S}}}$ and
$\delta_{{\phi_y}_{\mbox{\scriptsize S}}}$ with $P_x$ and
$P_{\phi_y}$, respectively, have 80~$\mu$m and 50~$\mu$rad accuracy.
The segment measurements were applied to correct the internal geometry
of the chambers.

\section{Alignment of CSC Chambers in Rings}
\label{sec:localcsc}

The CSC chambers in the muon endcaps were designed to overlap slightly
along the edges of the sensitive area, such that muons passing through
the narrow ``overlap regions'' would be observed by both of the
neighboring chambers.  This allows for a high-precision relative
alignment of neighbors, and the relative measurements can be propagated
around each ring of mutually overlapping chambers (illustrated in
Fig.~\ref{fig:muoneverything}).  All endcap rings are internally
connected in this way except ME$\pm$1/3.

Although CSC chamber alignment using overlap tracks and alignment with tracks from the tracker (described
in the next section) both determine the relative positions of CSC
chambers, the overlap method has two advantages: (1) it achieves high
precision with a small number of tracks, and (2) it is less prone to
potential systematic errors in tracking.  Since the two methods use
different datasets in different ways, comparison between them can be
used to diagnose systematic errors in tracks from the tracker or in their propagation
to the muon endcaps.  The disadvantage of the overlap alignment method
is that it does not relate the ring coordinate frame to the AIF.  A
second alignment step is therefore necessary to align the whole ring
as a rigid body relative to tracks from the tracker.

This section begins with a mathematical derivation of the algorithm,
which is an analytic solution of Eq.~(\ref{eqn:chi2millepede}) without
external constraints, followed by an analysis of results from a Monte
Carlo (MC) simulation and LHC single-beam tests.  Cosmic ray results
are not presented because the procedure requires forward-pointing
tracks with approximate azimuthal symmetry.

\subsection{Ring Alignment Algorithm}

The basic strategy of CSC ring alignment is to fit 
segments from the same muon in each of the overlapping chambers
independently and require them to match in position and slope for all
chambers in the ring simultaneously.  Segment parameters
$\overline{r\phi}$ and $\frac{\textrm{d}(r\phi)}{\textrm{d}z}$ are computed using only
cathode strips (the high-precision $r\phi$ coordinate), in a
coordinate frame shared by pairs of chambers with $z=0$ being the
plane of symmetry between them.  The curvature added to the segments
by parameterizing them with a curvilinear coordinate is negligible.
Chambers are labeled by indices $i$ ranging from $1$ to $N$ ($N=18$ or
$36$, depending on the ring), with the convention that $N+1$ refers to
chamber $1$.

The general alignment objective function (see
Eq.~(\ref{eqn:chi2millepede}) for definitions) could be applied here,
removing survey constraints and considering each track to consist of
two ``hits'' with two components each: $\overline{r\phi}^i_j$ and
$\frac{\textrm{d}(r\phi)}{\textrm{d}z}^i_j$.  However, the geometry of this
alignment case allows for simplifications which make the solution
analytic.  Overlap tracks only connect neighboring chambers, $i$ and
$i+1$, in such a way that residuals $\Delta (r\phi)^i_j
= \overline{(r\phi)}^i_j - \overline{(r\phi)}^{i+1}_j$ and
$\Delta \frac{\textrm{d}(r\phi)}{\textrm{d}z}^i_j = \frac{\textrm{d}(r\phi)}{\textrm{d}z}^i_j
- \frac{\textrm{d}(r\phi)}{\textrm{d}z}^{i+1}_j$ constrain
$A_j \cdot \left(\vec{\delta}^i - \vec{\delta}^{i+1}\right)$.
Minimization of terms with the form
\begin{equation}
\left(\begin{array}{c}
\Delta (r\phi)^i_j \\
\Delta \frac{\textrm{d}(r\phi)}{\textrm{d}z}^i_j \\
\end{array} \right) - A_j \cdot \left(\vec{\delta}^i - \vec{\delta}^{i+1}\right)
\end{equation}
is functionally equivalent to Eq.~(\ref{eqn:chi2millepede}) with
$B_i \cdot \delta \vec{p}_j$ explicitly evaluated.

The overlap region is narrow in $x$ and $\frac{\textrm{d}x}{\textrm{d}z}$ but wide in
$y$, so three alignment parameters, $\delta_{r\phi}$,
$\delta_{\phi_y}$, $\delta_{\phi_z}$ can be accessed from these data
(illustrated in Fig.~\ref{fig:coordinates-b}).  The expression
$A_j \cdot \left(\vec{\delta}^i - \vec{\delta}^{i+1}\right)$ expands to
\begin{equation}
\left(\begin{array}{c c c}
1 & 0 & -y_j \\
0 & 1 & 0 \\
\end{array}\right)
\left(\begin{array}{c}
\delta_{r\phi}^i - \delta_{r\phi}^{i+1} \\
\delta_{\phi_y}^i - \delta_{\phi_y}^{i+1} \\
\delta_{\phi_z}^i - \delta_{\phi_z}^{i+1} \\
\end{array}\right) \quad\mbox{.}
\end{equation}

Furthermore, correlations between $\delta_{\phi_y}^i
- \delta_{\phi_y}^{i+1}$ and $\delta_{r\phi}^i - \delta_{r\phi}^{i+1}$
can be replaced with an order dependence.  Rotating chambers (with
segments following their orientation) to make segments parallel
($\phi_y$ alignment) would change their intercepts at $z=0$,
but translating chambers to make segments continuous at $z=0$ ($r\phi$
alignment) would not change their slopes.  If
$\phi_y$ is aligned first and all segments are refitted, recomputing all residuals
with the new geometry, a subsequent alignment of $r\phi$ (and $\phi_z$) does not disturb
the $\phi_y$ minimization.  Therefore, two objective
functions, ${\chi_1}^2$ and ${\chi_2}^2$, can be minimized separately as long as
${\chi_1}^2$ is optimized first, and the derived geometry is used to calculate quantities in ${\chi_2}^2$.

Putting all of this together, 
\begin{eqnarray}
{\chi_1}^2 &=& \sum_i^{\mbox{\scriptsize chambers}} \sum_j^{\mbox{\scriptsize tracks}}
\left[ \Delta \frac{\textrm{d}(r\phi)}{\textrm{d}z}^i_j - \left(\delta_{\phi_y}^i
- \delta_{\phi_y}^{i+1}\right) \right]^2 \frac{1}{({\sigma_{\frac{\textrm{d}(r\phi)}{\textrm{d}z}}}^2)^i_j} \\
{\chi_2}^2 &=& \sum_i^{\mbox{\scriptsize chambers}} \sum_j^{\mbox{\scriptsize tracks}}
\left[ \Delta (r\phi)^i_j - \left(\delta_{r\phi}^i - \delta_{r\phi}^{i+1}\right)
+ y_j \left(\delta_{\phi_z}^i - \delta_{\phi_z}^{i+1}\right) \right]^2 \frac{1}{({\sigma_{r\phi}}^2)^i_j} \quad\mbox{,} \nonumber
\end{eqnarray}
where $({\sigma_{\frac{\textrm{d}(r\phi)}{\textrm{d}z}}}^2)^i_j$ and $({\sigma_{r\phi}}^2)^i_j$ are
one-parameter errors in the residuals.

The sum over tracks in ${\chi_1}^2$ can be recognized as a weighted
mean and the sum over tracks in ${\chi_2}^2$ as a linear fit of
$\Delta (r\phi)^i_j$ versus $y$.  Evaluating them (and introducing
$m$, $a$, and $b$ as functions returning the weighted mean, $y$
intercept, and slope versus $y$ of a given dataset, respectively), the objective functions can be replaced with
\begin{eqnarray}
{\chi_1'}^2 &=& \sum_i^{\mbox{\scriptsize chambers}}
\left[m\big(\big\{\Delta \tfrac{\textrm{d}(r\phi)}{\textrm{d}z}^i_j\big\}\big)
- \left(\delta_{\phi_y}^i - \delta_{\phi_y}^{i+1}\right)\right]^2 \\
{\chi_{2a}'}^2 &=& \sum_i^{\mbox{\scriptsize chambers}}
\left[a\big(\big\{\Delta (r\phi)^i_j, \, y_j\big\}\big) - \left(\delta_{r\phi}^i
- \delta_{r\phi}^{i+1}\right)\right]^2 \nonumber \\
{\chi_{2b}'}^2 &=& \sum_i^{\mbox{\scriptsize chambers}}
\left[b\big(\big\{\Delta (r\phi)^i_j, \, y_j\big\}\big) - \left(\delta_{\phi_z}^i - \delta_{\phi_z}^{i+1}\right)\right]^2\quad\mbox{,} \nonumber
\end{eqnarray}
where ${\chi_1'}^2$ differs from ${\chi_1}^2$ by a constant factor and
${\chi_{2a}'}^2 + {\chi_{2b}'}^2$ differs from ${\chi_2}^2$ by a
constant factor.  To avoid rotations and twists of the whole ring, Lagrange multiplier
$(\sum_i \delta_{\phi_y}^i/N)^2$ is added to ${\chi_1'}^2$,
$(\sum_i \delta_{r\phi}^i/N)^2$ is added to ${\chi_{2a}'}^2$, and
$(\sum_i \delta_{\phi_z}^i/N)^2$ is added to ${\chi_{2b}'}^2$,
which favor solutions with minimal average corrections.  Each
of the three minimization problems has the same analytic solution,
found by setting the derivatives of the objective function to zero and
inverting the resulting $N\times N$ matrix of constants.  This
alignment method is applicable to any tracking system composed of a
ring of pairwise overlapping detectors.

The method also enables three internal cross-checks; the following
must be consistent with zero: $\sum_i a(\{\Delta (r\phi)^i_j, \, y_j\})$, $\sum_i
b(\{\Delta (r\phi)^i_j, \, y_j\})$, and $\sum_i m(\{\Delta \frac{\textrm{d}(r\phi)}{\textrm{d}z}^i_j\})$.  These
closure tests are not sensitive to misalignments, but they determine
whether the residuals are consistent with a closed loop.  All three
closure tests were found to be consistent with zero in these studies.

\subsection{Monte Carlo Study}

The procedure was applied to beam-halo events generated by Monte Carlo
(based on a simulation described in Ref.~\cite{CMS_NOTE_2005-012}).  The
simulation has approximately the same number of events as the 2008 LHC
dataset (33$\,$000 tracks passing through the overlap regions of
CSCs), but a different azimuthal and radial distribution.  The
distribution of beam-halo events is difficult to predict for a new
accelerator, and in fact changed frequently during the first
circulating beam tests.  Since the alignment uncertainties are
statistics-limited and the population of tracks in each chamber is
only approximately the same as in data, the simulation provides only a
rough guide for what to expect from the data.

From an initially misaligned detector, the procedure aligned
$\delta_{\phi_y}$ with 1.04~mrad accuracy (initially 2~mrad), $\delta_{r\phi}$ with
230~$\mu$m (initially 1000~$\mu$m), and $\delta_{\phi_z}$ with
0.25~mrad (initially 1~mrad), determined from the
RMS of differences between the aligned positions and the true
positions of the chambers.

The second step, aligning the internally aligned rings relative to the
tracker, was studied with a sample of simulated muons from proton
collisions.  It was found that 280~$\mu$m ring position accuracy in
the $X$-$Y$ plane can be achieved with 10~pb$^{-1}$ of collisions.
This second step cannot be applied with beam-halo muons because they
generally do not cross both the muon chambers and the tracker.

\subsection{Alignment Results}

In September 2008, protons circulated in the LHC, producing beam-halo
muons captured by CMS.  The majority of the beam-halo data were
collected from one 9-minute fill of the anti-clockwise beam.  More
beam-halo muons illuminated the inner rings of the negative endcap
because they tend to follow trajectories close to the beam-line and
the anti-clockwise beam traverses CMS from the negative side.  All
chambers in ME$-$2/1 and ME$-$3/1 were operational, so these chambers
were used to test the procedure.

Application of the alignment algorithm narrows the RMS of the $\Delta
(r\phi)$ distribution from 1.42 to 0.98~mm.  For comparison, the
$\Delta (r\phi)$ distribution of the aligned simulation has an RMS of
1.12~mm.

To independently verify the results, they can be compared with
photogrammetry measurements.  The resolution in the positions of
photogrammetry targets is 300~$\mu$m~\cite{ref:endcapPG}, which
translates into resolutions of 210~$\mu$m and 0.23~mrad for
$\delta_{r\phi}$ and $\delta_{\phi_z}$ respectively.  Photogrammetry measurements cannot
determine $\delta_{\phi_y}$ because the targets lie near the $y$ axis
of the chambers.

\begin{figure}
\centering
\includegraphics[width=0.75\linewidth]{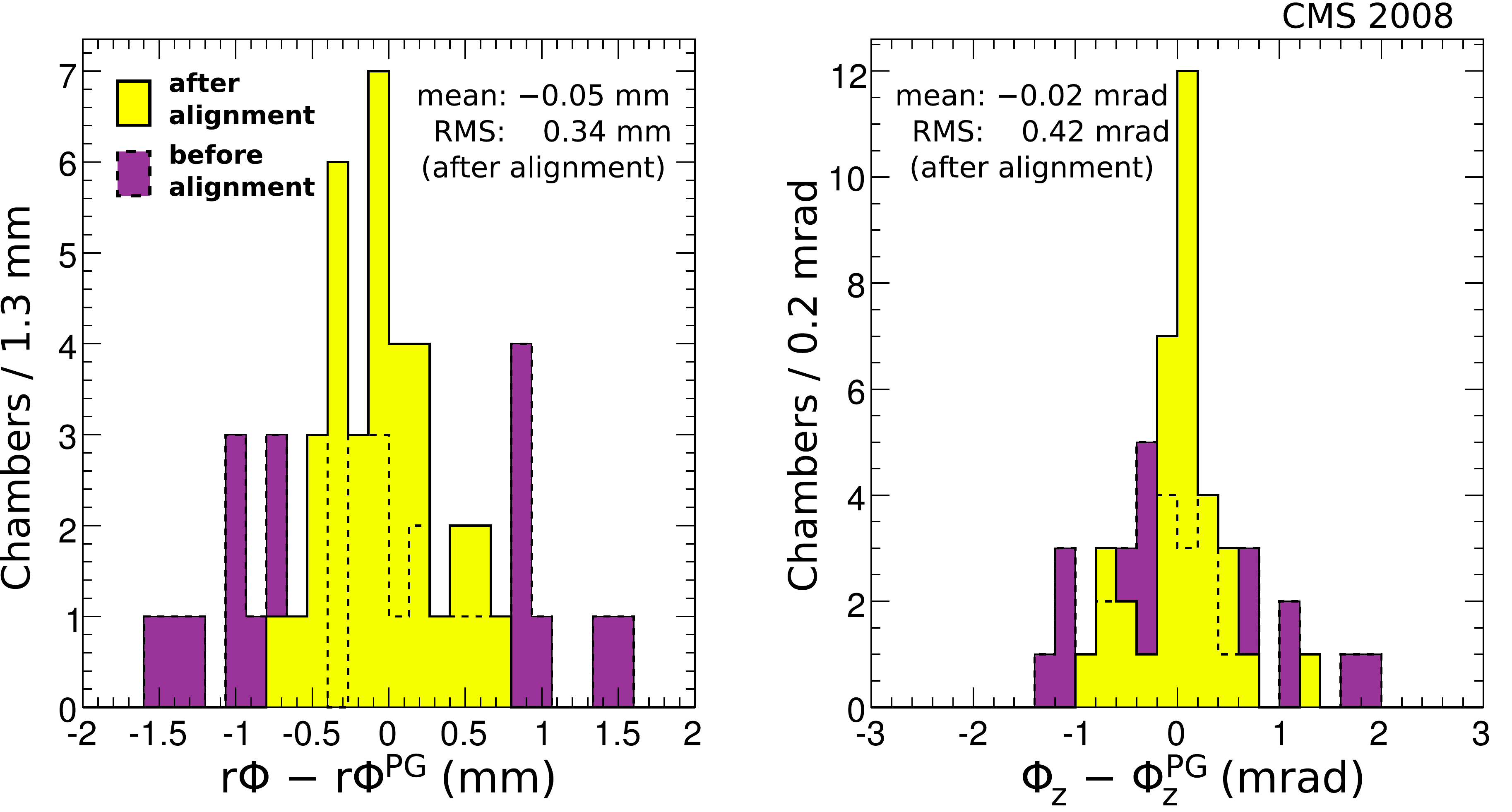}
\caption{Difference between CSC positions and their photogrammetry
measurements.  Dark/dashed histograms are before alignment
and light histograms are after alignment.  \label{fig:overlaps_data2}}
\end{figure}

Figure~\ref{fig:overlaps_data2} shows the difference between chamber
coordinates and photogrammetry (${r\phi}^{\mbox{\scriptsize PG}}$ and
${\phi_z}^{\mbox{\scriptsize PG}}$), before and after alignment.  The
RMS of these distributions after alignment, which are convolutions of 
photogrammetry errors and errors in the track-based measurements, are
340~$\mu$m and 0.42~mrad in $\delta_{r\phi}$ and $\delta_{\phi_z}$ respectively.
Subtracting the photogrammetry errors in quadrature, one can conclude that
the track-based measurement has approximately 270~$\mu$m and 0.35~mrad
uncertainties, in rough agreement with the prediction from simulation.
In the absence of systematic uncertainties, several hours of similar
beam-halo conditions would be sufficient to reduce the alignment error
below the 170--200~$\mu$m intrinsic hit uncertainty for these
chambers~\cite{ref:csc_resolution}.

\section{Alignment of DT Chambers in a Global Coordinate System}
\label{sec:global_muon_alignment}

The muon alignment procedures presented in previous sections arrange
layers and sets of chambers in self-consistent coordinate frames, but
do not relate those frames to the other subsystems of CMS, in
particular the tracker.  In this section, a method is described to
align muon chambers relative to the tracker by propagating the tracker
tracks to the muon chambers.  The method is equally applicable to DTs
and CSCs, but cosmic rays only provide large numbers of tracks in the
central region of the barrel.

Muons encounter significant scattering material between every two
stations.  With measurements expressed as chamber residuals
($\Delta x$, $\Delta y$, $\Delta \frac{\textrm{d}x}{\textrm{d}z}$,
$\Delta \frac{\textrm{d}y}{\textrm{d}z}$), this means there is a large component of
random error in the trajectory between each measurement and the next.
The residuals are therefore broadened beyond what would be expected
from the intrinsic resolutions of the hits, but alignment information
can be derived from the peak positions of those distributions.

In principle, muon chamber hits could be used in the track fits to
narrow the distributions of residuals and improve the statistical
precision of the alignment results.  Including those hits biases the tracks,
coupling track parameters and alignment parameters, thereby coupling
alignment parameters of different chambers with each other.  This coupling can be
resolved by matrix inversion~\cite{Blobel:2006yh} or by reducing the
weight of the muon chamber hits and iterating~\cite{Karimaki:2003bd}, but
the structure of the probability distribution from scattering
complicates these methods.  Moreover, sufficient statistical precision
can be achieved without muon chamber hits in the track fits (or
equivalently, muon chamber hits with negligible weight in the fit).

This section describes muon chamber alignment using tracks
determined purely by the tracker.  Because the track parameters are
not a function of the muon chamber positions, there is no mutual
dependency to resolve, nor are the alignment parameters of different
chambers coupled.  However, the resulting chamber positions depend on
the alignment of the tracker~\cite{Collaboration:2009sr}: optimized
muon chamber residuals does not guarantee that the combined
tracker-muon system is globally undistorted.  The first subsection
below describes the algorithm with a discussion of propagation
effects.  Section~\ref{sec:gmaresults} presents a Monte Carlo study
of the procedure, and alignment results are presented in
Sections~\ref{sec:gmaresults2} (residuals),
\ref{sec:gmavalidation} (cross-check), and \ref{sec:momentum} (momentum measurements).
An alternative algorithm is under development to align muon chambers
using matrix inversion; this is briefly described in
Section~\ref{sec:mpalgo}.

\subsection{The Reference-Target Algorithm}
\label{sec:rtalgo}

The ``reference-target'' algorithm divides the tracking volume into two
regions: a ``reference'' (the tracker), in which normal track-fitting is
performed, and a ``target'' (the muon chambers), in which unbiased
residuals are computed from the propagated tracks.  The simplicity
this affords in the correlation matrix of alignment parameters allows
more emphasis to be placed on the study of propagation effects.

The shape of the residuals distribution is fitted to a parameterized
Ansatz function using an unbinned maximum likelihood method, rather than
minimizing an objective function with a quadratic form like
Eq.~(\ref{eqn:chi2millepede}).  This can be seen as a
generalization of the standard method, because quadratic terms such as
$(\Delta x_{ij} -
A_j \cdot \vec{\delta}_i)^2/({\sigma_{\mbox{\scriptsize hit}}}^2)_{ij}$
are the logarithm of Gaussian likelihoods.  Substituting a
non-Gaussian Ansatz motivated by physical processes in track
propagation introduces non-linearity to the derivative of the
objective function which cannot be solved by matrix inversion.  
The non-linear minimization package MINUIT is used instead~\cite{James:1975dr}.

\subsubsection{Propagation Effects}

Residuals from propagated tracks can be affected by the following effects:
\begin{itemize}
\item misalignment (distributed as a $\delta$-function in $x$, $y$,
$\frac{\textrm{d}x}{\textrm{d}z}$, $\frac{\textrm{d}y}{\textrm{d}z}$ because it
is \mbox{strictly geometric);\hspace{-0.5 cm}}
\item intrinsic hit resolution (negligible);
\item statistical uncertainty in the fitted track parameters (Gaussian);
\item hard Coulomb scattering off of nuclei (which has non-Gaussian tails);
\item multiple Coulomb scattering (Gaussian in the limit of many interactions);
\item background from pattern-recognition errors and
noisy channels (non-Gaussian);
\item systematic errors in magnetic field map and material budget for
average energy loss (proportional to $q/p_T$ and $q/|\vec{p}|^2$,
respectively, where $q$, $p_T$, and $|\vec{p}|$ are the charge,
transverse momentum, and magnitude of the momentum of the muon);
\item systematic bias in the track source distribution (a function of
the path of the track through the reference volume).
\end{itemize}
All but the last two effects are included in the Ansatz.  The main
non-Gaussian contribution is from events in which the muon interacts
with a small number of nuclei, such that the distribution of
deflections does not fully reach the Gaussian limit of the central
limit theorem.  There is no sharp distinction between single and
multiple scattering, but it is sufficient to model the combined effect
with a function having a Gaussian core and large tails, such as the
tails of a Cauchy-Lorentz distribution.  A smaller non-Gaussian
contribution from pattern-recognition errors and noisy channels is
also absorbed into the tails.  The primary significance of the tails
is to increase the likelihood of highly non-Gaussian residuals, and
therefore stabilize the determination of the peak.

A magnetic field map resulting from detailed modeling and CRAFT data
analysis~\cite{ref:magnetic_field} was used in this alignment, but the
result is additionally verified by performing it separately with
positively charged muons, negatively charged muons, and averaging the
two.  Magnetic field errors add contributions to residuals which are
antisymmetric in charge, as do material budget errors because muons
can only lose energy on average, and this deflects them in the
direction of their curvature.  Any charge-dependent effects from
magnetic field map or material budget errors would cancel in the
average.  The RMS of differences with respect to not applying this
averaging procedure are 100~$\mu$m in $\delta_x$ and 0.07~mrad in
$\delta_{\phi_y}$, the two parameters most affected by magnetic field.
These differences are small compared with other uncertainties.

The possibility of systematic bias in the track source has been
addressed in the context of tracker
alignment~\cite{Flucke:2008zz,Stoye:1047047} as weak modes in the
procedure and has been studied in Ref.~\cite{Collaboration:2009sr} for
the tracker description used here.  Any unresolved global
distortions in the tracker would be extended to the muon system as
well, though tracks would be guaranteed to match segments in the
momentum range of the algorithm's application.  The goal at this stage
is to align muon chambers to the positions projected by the current
best knowledge of the shape of the tracker.

\subsubsection{The Alignment Ansatz Function}

Misalignment offsets the peak of the residuals distribution, centering
($\Delta x$, $\Delta y$, $\Delta \frac{\textrm{d}x}{\textrm{d}z}$,
$\Delta \frac{\textrm{d}y}{\textrm{d}z}$) at the values
\begin{equation}
\renewcommand{\arraystretch}{2.5}
\left(\begin{array}{c}
{\Delta x}_0 \\
{\Delta y}_0 \\
{\Delta \dfrac{\textrm{d}x}{\textrm{d}z}}_0 \\
{\Delta \dfrac{\textrm{d}y}{\textrm{d}z}}_0 \\
\end{array}\right)
=
{\renewcommand{\arraystretch}{2.5}
\left(\begin{array}{c c c c c c}
1 & 0 & -\dfrac{\textrm{d}x}{\textrm{d}z} & -y \dfrac{\textrm{d}x}{\textrm{d}z} & x \dfrac{\textrm{d}x}{\textrm{d}z} & -y \\
0 & 1 & -\dfrac{\textrm{d}y}{\textrm{d}z} & -y \dfrac{\textrm{d}y}{\textrm{d}z} & x \dfrac{\textrm{d}y}{\textrm{d}z} & x \\
0 & 0 & 0 & -\dfrac{\textrm{d}x}{\textrm{d}z} \dfrac{\textrm{d}y}{\textrm{d}z} & 1 + \left(\dfrac{\textrm{d}x}{\textrm{d}z}\right)^2 & -\dfrac{\textrm{d}y}{\textrm{d}z} \\
0 & 0 & 0 & -1 - \left(\dfrac{\textrm{d}y}{\textrm{d}z}\right)^2 & \dfrac{\textrm{d}x}{\textrm{d}z}\dfrac{\textrm{d}y}{\textrm{d}z} & \dfrac{\textrm{d}x}{\textrm{d}z}
\end{array}\right)}
\renewcommand{\arraystretch}{1.7}
\left(\begin{array}{c}
\delta_x \\
\delta_y \\
\delta_z \\
\delta_{\phi_x} \\
\delta_{\phi_y} \\
\delta_{\phi_z}
\end{array}\right) \quad\mbox{,}
\label{eqn:dtmatrix}
\end{equation}
where $(x,y)$ represents the coordinates of the track intersection
with the chamber and $(\frac{\textrm{d}x}{\textrm{d}z},\frac{\textrm{d}y}{\textrm{d}z})$ the entrance
angle.  The above matrix is an extension of Eq.~(17) in
Ref.~\cite{Karimaki:2003bd} to include angular residuals
$\Delta \frac{\textrm{d}x}{\textrm{d}z}$ and $\Delta \frac{\textrm{d}y}{\textrm{d}z}$, significantly
increasing sensitivity to $\delta_{\phi_y}$ and $\delta_{\phi_x}$,
respectively.

A Voigt distribution, or convolution of a Gaussian with a
Cauchy-Lorentzian, is used to model the Gaussian core with large tails.
The function
\begin{equation}
f(t; \, t_0, \, \sigma, \, \Gamma) = \int_{-\infty}^\infty
\frac{1}{\pi}\frac{\Gamma/2}{(t - s - t_0)^2 + (\Gamma/2)^2} \times
\frac{1}{\sqrt{2\pi} \sigma} \exp\left(\frac{-s^2}{2
  \sigma^2}\right) \, ds\quad\mbox{,}
\label{eqn:fitfunction}
\end{equation}
has one variable $t$ with three parameters $t_0$, $\sigma$,
and $\Gamma$.  Close to the peak, the distribution is approximately
Gaussian (because $\Gamma \ll \sigma$, typically by a factor of 10), and far from the
peak, the distribution is approximately $1/t^2$.

The fit function for the four-dimensional residuals distribution is
built from a product of four Voigt distributions.  To account for
correlations between $\Delta x$ and $\Delta \frac{\textrm{d}x}{\textrm{d}z}$, and
between $\Delta y$ and $\Delta \frac{\textrm{d}y}{\textrm{d}z}$, parameters
$\alpha_{\Delta x}$ and $\alpha_{\Delta y}$ express linear dependences between
them in the fit function, and are allowed to float freely in the alignment fit.
The correlation is simply due to the fact that an
error in the track direction $\Delta \frac{\textrm{d}x}{\textrm{d}z}$ introduced at a
distance $L$ upstream of the chamber causes a $\Delta x \approx
L \Delta \frac{\textrm{d}x}{\textrm{d}z}$ error in position.

\begin{figure}
\begin{center}
\includegraphics[height=\linewidth, angle=90]{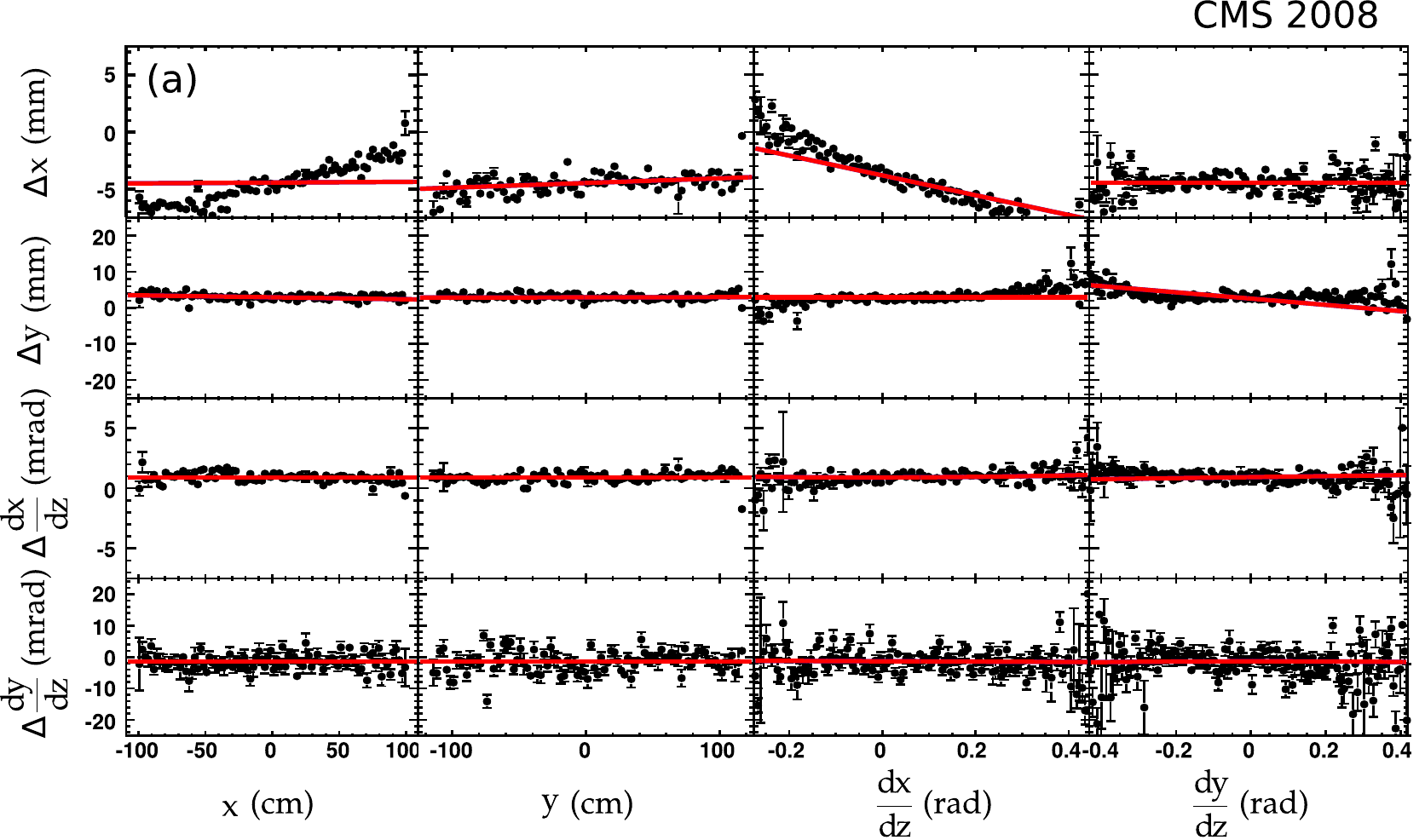}

\includegraphics[height=\linewidth, angle=90]{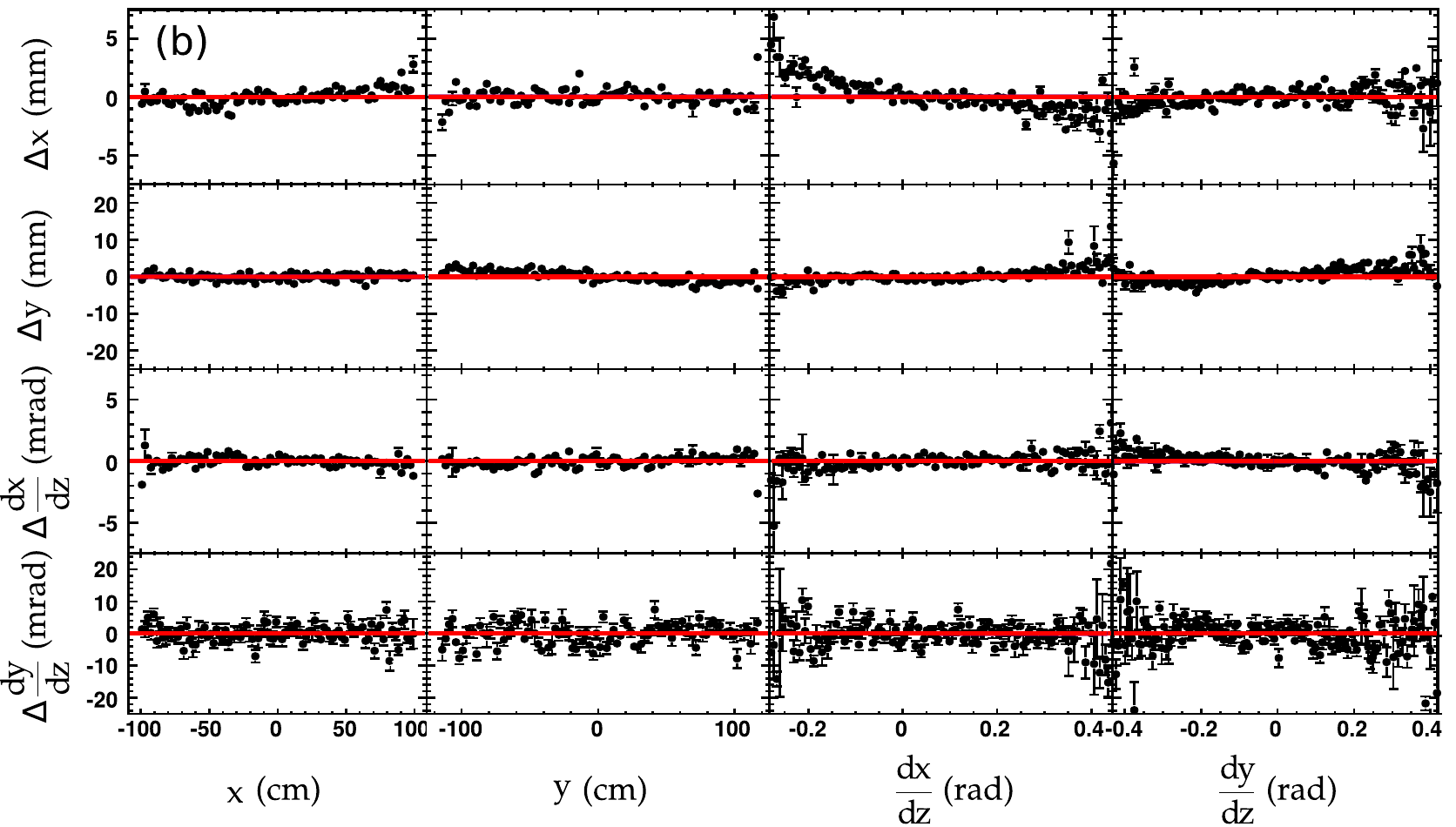}
\end{center}

\caption{Mean residuals with statistical error bars versus position ($x$,~$y$)
and entrance angle ($\frac{\textrm{d}x}{\textrm{d}z}$,~$\frac{\textrm{d}y}{\textrm{d}z}$) in one chamber
(DT wheel~0, station~1, sector~10) from cosmic-ray data, (a) before
alignment and (b) after alignment.  Each bin selects a narrow range of
the position or entrance angle component under consideration, but
averages over all other variables.  The fit function, evaluated at
zero in all other variables, is overlaid as lines before and after
alignment.  Asymmetries in the position and entrance angle
distributions allow for misalignments to be manifested in more ways in
the averaged residuals than in the projections of the best-fit function, but the fit properly
removes these misalignments, resulting in nearly flat lines in all variables.
\label{fig:examplefit}}
\end{figure}

Without explicitly substituting the alignment parameters, the full fit
function is
\begin{multline}
F\bigg(\Delta x, \Delta y, \Delta \tfrac{\textrm{d}x}{\textrm{d}z}, \Delta \tfrac{\textrm{d}y}{\textrm{d}z}; \\
{\Delta x}_0, {\Delta y}_0, {\Delta \frac{\textrm{d}x}{\textrm{d}z}}_0, {\Delta \frac{\textrm{d}y}{\textrm{d}z}}_0,
\sigma_{\Delta x}, \sigma_{\Delta y}, \sigma_{\Delta \tfrac{\textrm{d}x}{\textrm{d}z}}, \sigma_{\Delta \tfrac{\textrm{d}y}{\textrm{d}z}},
\Gamma_{\Delta x}, \Gamma_{\Delta y}, \Gamma_{\Delta \tfrac{\textrm{d}x}{\textrm{d}z}}, \Gamma_{\Delta \tfrac{\textrm{d}y}{\textrm{d}z}},
\alpha_{\Delta x}, \alpha_{\Delta y}) = \\
f\bigg(\Delta x; \, \big({\Delta x}_0 + \alpha_{\Delta x} \Delta \tfrac{\textrm{d}x}{\textrm{d}z}\big), \, \sigma_{\Delta x}, \, \Gamma_{\Delta x}\bigg) \, \times \,
f\bigg(\Delta \tfrac{\textrm{d}x}{\textrm{d}z}; \, {\Delta \tfrac{\textrm{d}x}{\textrm{d}z}}_0, \, \sigma_{\Delta \tfrac{\textrm{d}x}{\textrm{d}z}}, \, \Gamma_{\Delta \tfrac{\textrm{d}x}{\textrm{d}z}}\bigg) \\
f\bigg(\Delta y; \, \big({\Delta y}_0 + \alpha_{\Delta y} \Delta \tfrac{\textrm{d}y}{\textrm{d}z}\big), \, \sigma_{\Delta y}, \, \Gamma_{\Delta y}\bigg) \, \times \,
f\bigg(\Delta \tfrac{\textrm{d}y}{\textrm{d}z}; \, {\Delta \tfrac{\textrm{d}y}{\textrm{d}z}}_0, \, \sigma_{\Delta \tfrac{\textrm{d}y}{\textrm{d}z}}, \, \Gamma_{\Delta \tfrac{\textrm{d}y}{\textrm{d}z}}\bigg)\quad\mbox{.}
\label{eqn:fitfunc}
\end{multline}
Substituting $\Delta x_0$, $\Delta y_0$, $\Delta {\frac{\textrm{d}x}{\textrm{d}z}}_0$, $\Delta {\frac{\textrm{d}y}{\textrm{d}z}}_0$
from Eq.~(\ref{eqn:dtmatrix}) introduces the track position and
entrance angle as four new variables and the alignment parameters as
six new parameters.  All parameter values and estimates of their
uncertainties are evaluated by MINUIT, seeded by the truncated mean
and RMS of each distribution.  An example fit is shown in Fig.~\ref{fig:examplefit}.

The alignment fit is weighted to reduce the influence of poorly formed
segments.  The quality of a segment is quantified by
$\chi^2/\mbox{ndf}$, rather than uncertainties in its parameters, so
segments in the alignment fit are weighted by $w_i =
(\mbox{ndf}/\chi^2)_i/\sum_{i'} (\mbox{ndf}/\chi^2)_{i'}$ of the
segment fits.  The objective function minimized by the alignment is
$\chi^2 = \sum_i w_i \log F_i$ where $F_i$ is given by
Eq.~(\ref{eqn:fitfunc}) for each segment $i$.  To avoid domination of
the fit by a few of the most linear segments, which are not
necessarily from the best-determined tracks (unscattered muons), segments with the largest 1\% of
weights have been excluded.

To resolve unmodeled non-linearities in the residuals, the procedure
is applied twice, taking the output of the first iteration as an
initial geometry for the second.  No subsequent improvements are
observed in a third iteration or beyond.

\subsubsection{Configuration for Alignment with Cosmic-Ray Muons}
\label{sec:configuration}

The vertical distribution of cosmic rays underground imposed geometric
restrictions on the set of chambers that could be aligned with this
algorithm.  Only barrel wheels $-$1, 0, $+$1, excluding horizontal
sectors~1 and 7 (see Fig.~\ref{fig:muoneverything}), recorded a
sufficient number of muons that also crossed the barrel of the tracker
to perform an alignment.  In addition, the fits for four nearly
horizontal chambers (in wheel, station, sector ($-$1, 2, 8), ($+$1, 3,
8), ($-$1, 1, 12), and ($+$1, 2, 2)) failed to converge, all for
reasons related to the scarcity of horizontal cosmic ray muons.

For most chambers within the restricted set, however, cosmic-ray muons are
sufficiently abundant that the measurement is not statistics-limited.
It is therefore possible to apply a tight set of track quality
requirements, to control systematic errors:
\begin{itemize}
\item 100 $<$ $p_T$ $<$ 200~GeV/$c$ (nearly straight tracks; the upper
limit guarantees statistical independence from one of the cross-checks);
\item 12 out of 12 hits in DT chambers of stations~1--3, 8 out of 8 hits
in DT chambers of station~4;
\item at least 15 hits on the tracker track, with $\chi^2/\mbox{ndf} <
10$.
\end{itemize}
The cosmic-ray period included several on-off cycles of the magnetic
field, and the full field (3.8~T) periods were shown to result in
reproducible alignment parameters with the hardware
system~\cite{ref:hardware_alignment} and track residuals.  Only
datasets marked as acceptable for physics with full field from the CMS
``run registry'' \cite{ref:CRAFTGeneral} were used.  Of the
270~million cosmic-ray triggered events in CRAFT, the above
requirements select $100\,000$ tracks (the reduction is primarily due
the $p_T$ selection and the requirement that cosmic-ray trajectories
to pass through the tracker), yet statistical uncertainties are
typically only 120~$\mu$m in $\delta_x$.  Systematic errors, which may
amount to several hundred microns, dominate the alignment parameter
uncertainties, and hence it is better to select the highest-quality
muons.

DT chambers in stations~1--3 are aligned in all six degrees of
freedom, but chambers of station~4 are only aligned in $\delta_x$,
$\delta_{\phi_y}$, and $\delta_{\phi_z}$, as these are the most sensitive
alignment parameters without $\Delta y$ and $\Delta \frac{\textrm{d}y}{\textrm{d}z}$
residuals (recall that station~4 chambers have no $y$-measuring
superlayer).  Unaligned coordinates are not allowed to float in the
minimization.

\subsection{Monte Carlo Study}
\label{sec:gmaresults}

To test the alignment algorithm, a large sample of cosmic rays was
simulated (using \mbox{CMSCGEN}, described in Ref.~\cite{Biallass:2009ev}),
tracks were reconstructed with misaligned muon chambers (2~mm Gaussian
smearing in $x$, 4~mm in $y$ and $z$, 2~mrad in $\phi_x$, $\phi_y$,
and $\phi_z$), and the algorithm was applied to restore the original
alignment parameters using only the tracks.  To focus on the accuracy
of the algorithm itself, the tracker geometry, internal muon layer
geometry, magnetic field map, and material distribution were modeled
without errors (identical in simulation and reconstruction).  All
other detector effects were realistically modeled.

\begin{figure}[p]
\includegraphics[height=\linewidth, angle=90]{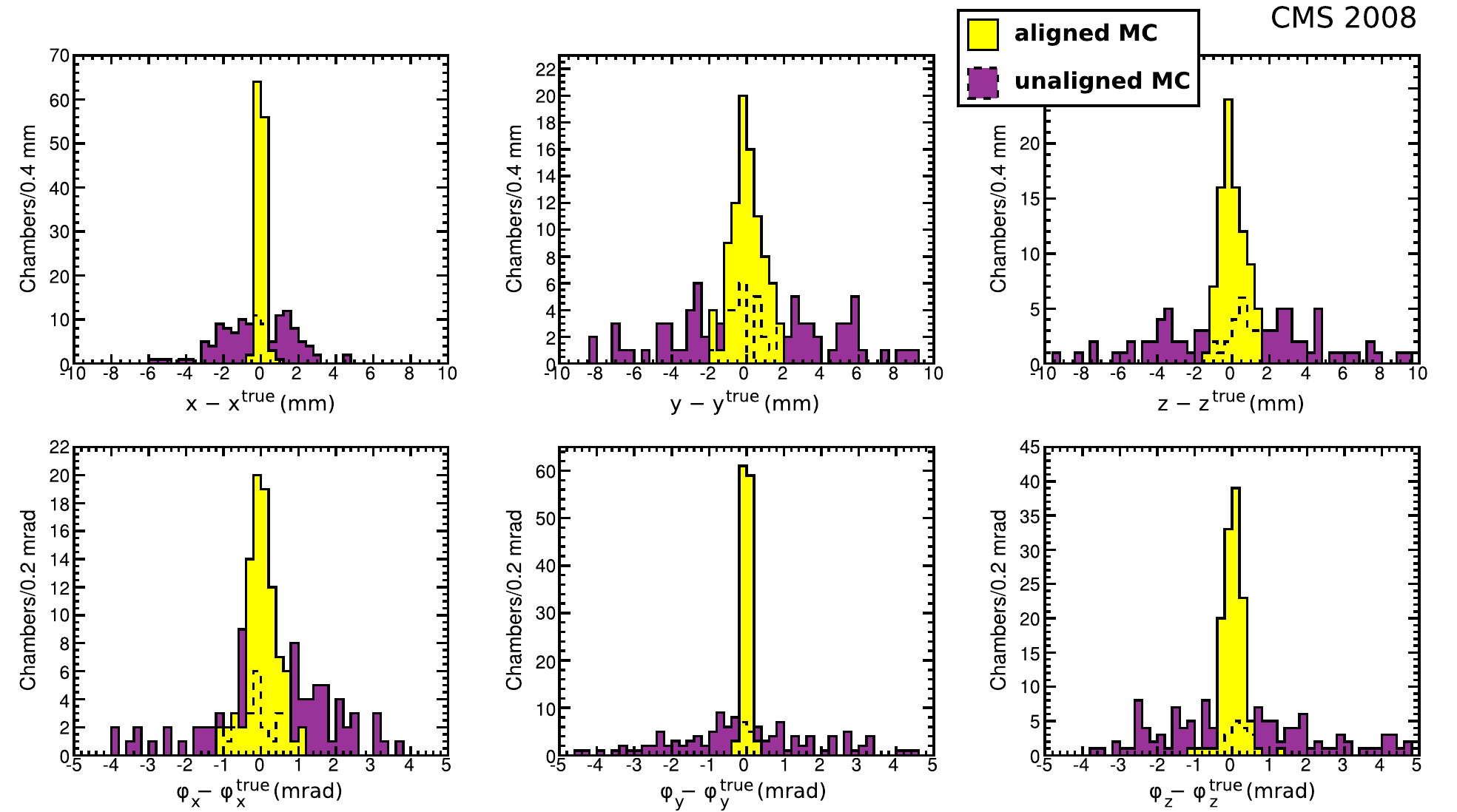}

\caption{Differences between reconstructed and true positions of muon
chambers from an alignment performed with $350\,000$ simulated cosmic-ray tracks passing selection requirements
(only showing chambers in wheels $-$1, 0, $+$1, all sectors except 1 and 7). \label{fig:hip_MC}}
\end{figure}

The test was performed twice, the first time with $350\,000$ tracks
passing all selection requirements and again with $100\,000$ tracks (a
subsample).  The first can be considered an infinite-statistics limit,
as statistical uncertainties from MINUIT are typically 3--4 times
smaller than the aligned position errors (see
Tables~\ref{tab:staterrs} and \ref{tab:hip_MC}).  The second matches
the statistical precision of CRAFT.  Distributions of chamber position
errors are presented in Fig.~\ref{fig:hip_MC}.

\begin{table}[p]
\caption{Statistical uncertainties in simulation and data:
uncertainties for all chambers $i=1$ to $N$ are summarized by
presenting $\sqrt{\frac{1}{N} \sum_i {\sigma_i}^2}$ where $\sigma_i$
is the statistical uncertainty in one of the six alignment parameters
below. \label{tab:staterrs}}
\begin{center}
\renewcommand{\arraystretch}{1.5}
\begin{tabular}{l | c c c c c c}
\hline\hline sample & $\delta_x$ (mm) & $\delta_y$ (mm) & $\delta_z$ (mm) & $\delta_{\phi_x}$ (mrad) & $\delta_{\phi_y}$ (mrad) & $\delta_{\phi_z}$ (mrad) \\\hline
$350$~k simulation & 0.059 & 0.118 & 0.248 & 0.170 & 0.038 & 0.072 \\\hline
$100$~k simulation & 0.106 & 0.210 & 0.443 & 0.305 & 0.069 & 0.129 \\
$100$~k data & 0.117 & 0.243 & 0.512 & 0.326 & 0.074 & 0.146 \\\hline\hline
\end{tabular}
\end{center}
\end{table}

\begin{table}[p]
\caption{RMS of differences between aligned and true positions of
chambers in Monte Carlo simulations (distributions in Fig.~\ref{fig:hip_MC}).  \label{tab:hip_MC}}
\renewcommand{\arraystretch}{1.5}
\begin{center}
\begin{tabular}{l | c c c c c c}
\hline\hline sample & $\delta_x$ (mm) & $\delta_y$ (mm) & $\delta_z$ (mm) & $\delta_{\phi_x}$ (mrad) & $\delta_{\phi_y}$ (mrad) & $\delta_{\phi_z}$ (mrad) \\\hline
$350$~k simulation & 0.192 & 0.841 & 0.630 & 0.417 & 0.095 & 0.287 \\
$100$~k simulation & 0.209 & 0.889 & 0.836 & 0.497 & 0.148 & 0.303 \\\hline\hline
\end{tabular}
\end{center}
\end{table}

\subsection{Alignment Results: Residuals Distribution}
\label{sec:gmaresults2}

\begin{figure}[p]
\centering
\includegraphics[height=0.9\linewidth, angle=90]{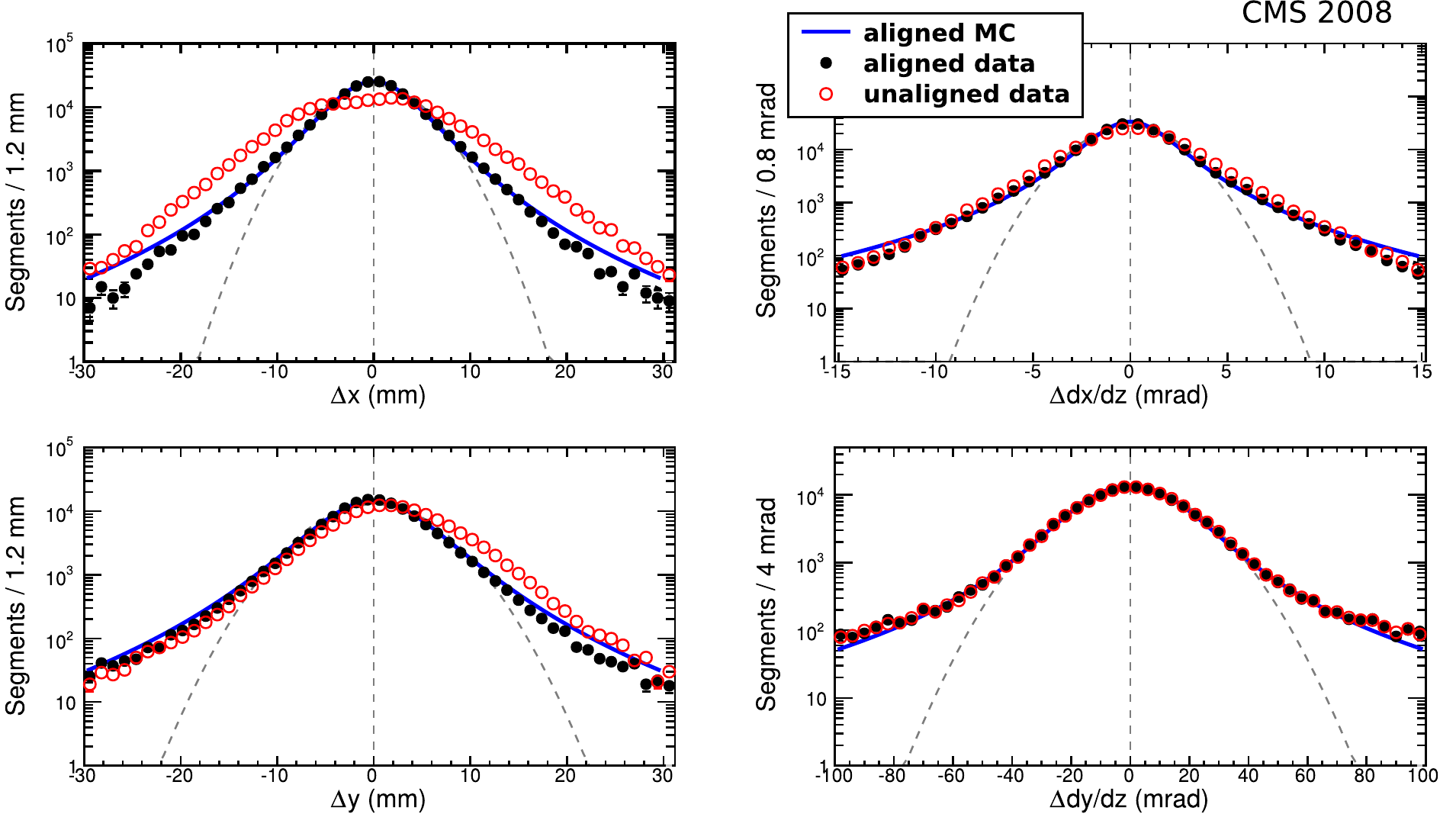}
\caption{Residuals distributions before and after alignment using
CRAFT data.  The full curves describe the aligned Monte
Carlo prediction and the light dashed curves indicate the peak
of each aligned distribution and its Gaussian approximation. \label{fig:residuals_all}}
\end{figure}

After applying the algorithm to the CRAFT cosmic-ray dataset,
the four-component distribution of residuals is found to be centered on zero
to the same degree as in the Monte Carlo simulation.
Figure~\ref{fig:residuals_all} presents the residuals distribution of
all aligned chambers, where the alignment makes the
distribution narrower and smoother.  The size of the dataset is also
large enough to see the non-Gaussian tails in detail, and that the
simulated residuals distribution closely matches the real one.
However, the raw residuals do not provide a sensitive probe of the
alignment accuracy, because alignment corrections are typically much
smaller than the width of the distribution.

\begin{figure}[p]
\centering
\includegraphics[height=0.9\linewidth, angle=90]{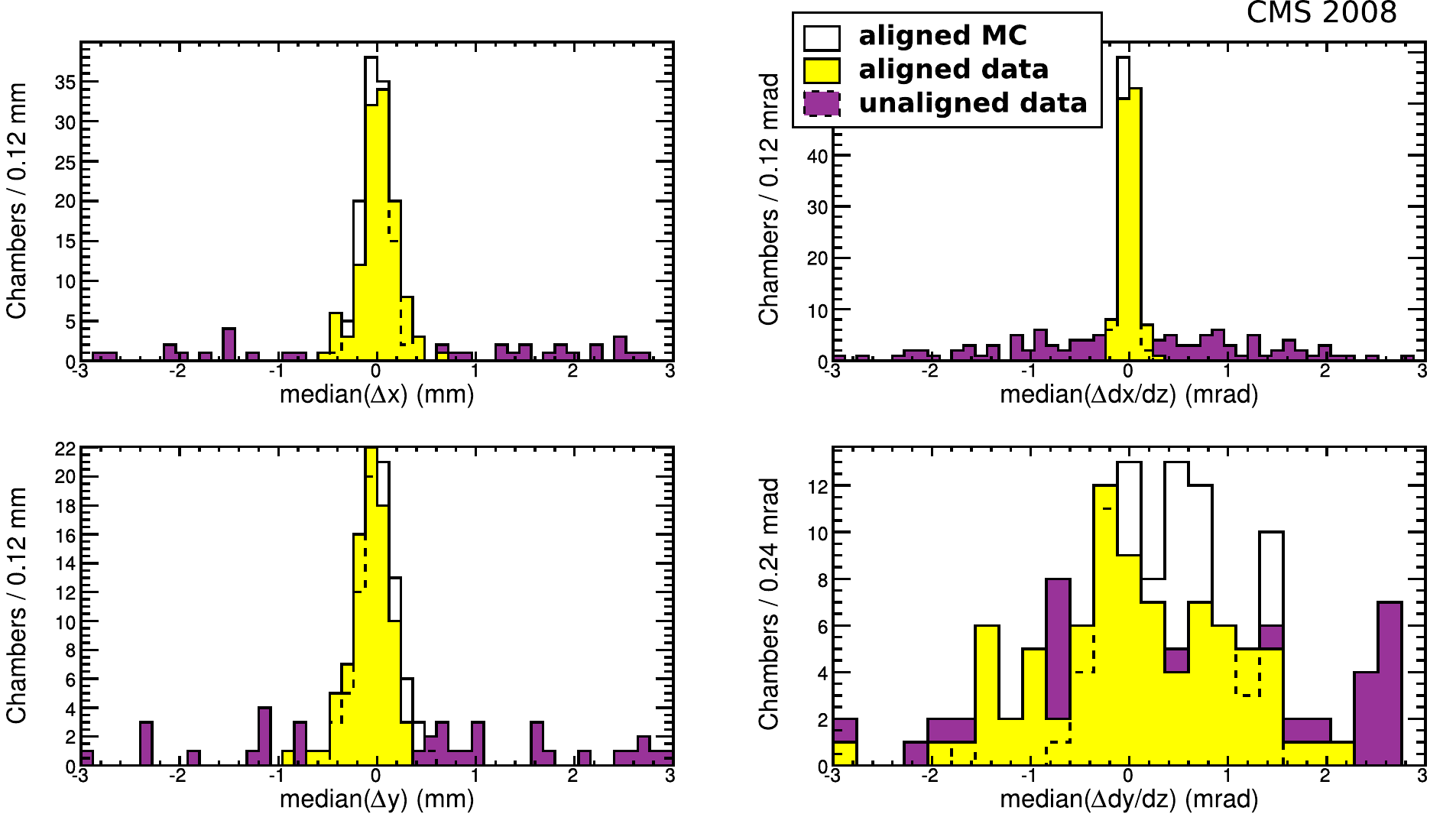}
\caption{Medians of residuals distributions by chamber (one histogram
entry per chamber). \label{fig:residuals_mean}}
\end{figure}

For a higher sensitivity, the median of the residuals
distribution of each chamber is calculated separately, then plotted as a distribution
in Fig.~\ref{fig:residuals_mean}, with the RMS of the distribution presented
in Table~\ref{tab:residuals_mean}.  The median is less affected by
non-Gaussian tails than the mean, and it is a different way of
achieving this insensitivity than the Voigt fits used by the
algorithm.

\begin{table}[p]
\caption{RMS of medians of residuals distributions by chamber (distributions in Fig.~\ref{fig:residuals_mean}).} \label{tab:residuals_mean}

\vspace{-0.5 cm}
\begin{center}
\renewcommand{\arraystretch}{1.5}
\begin{tabular}{l | c c c c}
\hline\hline & \hspace{0.3 cm}$x$ (mm)\hspace{0.3 cm} & \hspace{0.3 cm}$y$ (mm)\hspace{0.3 cm} & \hspace{0.3 cm}$\frac{\textrm{d}x}{\textrm{d}z}$ (mrad)\hspace{0.3 cm} & \hspace{0.3 cm}$\frac{\textrm{d}y}{\textrm{d}z}$ (mrad)\hspace{0.3 cm} \\\hline
Aligned MC ($100$~k) & 0.159 & 0.172 & 0.066 & 0.630 \\
Aligned data ($100$~k) & 0.190 & 0.166 & 0.085 & 0.885 \\\hline
Unaligned data ($100$~k) & 5.667 & 2.570 & 1.316 & 1.605 \\\hline\hline
\end{tabular}
\end{center}
\end{table}

\subsection{Alignment Results: Cross-check}
\label{sec:gmavalidation}

The analysis of residuals in the previous subsection provides
confidence that the alignment algorithm is operating as designed.  This section
presents a test of the aligned geometry using a
significantly different method, namely local segment fits.

A study of alignment parameter consistency for neighboring chambers
using overlapping segments from CSCs, as described in
Section~\ref{sec:localcsc}, would provide an ideal cross-check.  However, most
cosmic rays fall on the barrel, in which chambers do not overlap one another (with the
exception of several chambers in station~4).  The local alignment quality is
therefore checked by comparing local segments from DT chambers in neighboring stations,
propagated through only one layer of steel to the next station.

For each sector in a pair of neighboring stations (MB1$\to$MB2,
MB2$\to$MB3, and MB3$\to$MB4), segments are linearly propagated from the
inner chamber to the outer chamber and the parameters $x^{\mbox{\scriptsize local}}$,
${\frac{\textrm{d}x}{\textrm{d}z}}^{\mbox{\scriptsize local}}$ of the
propagated segment are compared with those of the segment in the outer chamber,
yielding two residuals, $\Delta x^{\mbox{\scriptsize local}}$,
$\Delta {\frac{\textrm{d}x}{\textrm{d}z}}^{\mbox{\scriptsize local}}$.

Curvature from the magnetic field is not included in this propagation,
but the error is corrected by taking advantage of the fact that
contributions to $\Delta x^{\mbox{\scriptsize local}}$ and $\Delta
{\frac{\textrm{d}x}{\textrm{d}z}}^{\mbox{\scriptsize local}}$ from the magnetic field
are charge-dependent.  Segments are associated with the corresponding
tracker tracks to identify their charge and to select high transverse momentum
($p_T$ $>$ 50~GeV/$c$).  Residuals from positively charged muons and
negatively charged muons are collected separately, fitted to Gaussian
distributions to identify the peaks, and averaged without weights.
Since the momentum spectra of positively and negatively charged
cosmic-ray muons are the same at this momentum scale, the magnetic
field contributions to average $\Delta x^{\mbox{\scriptsize local}}$
and $\Delta {\frac{\textrm{d}x}{\textrm{d}z}}^{\mbox{\scriptsize local}}$ cancel,
leaving only differences from misalignments.  This is the same
procedure as used to test sensitivity to the magnetic field by the
reference-target algorithm.

\begin{figure}[p]
  \centering
  \includegraphics[height=\linewidth, angle=90]{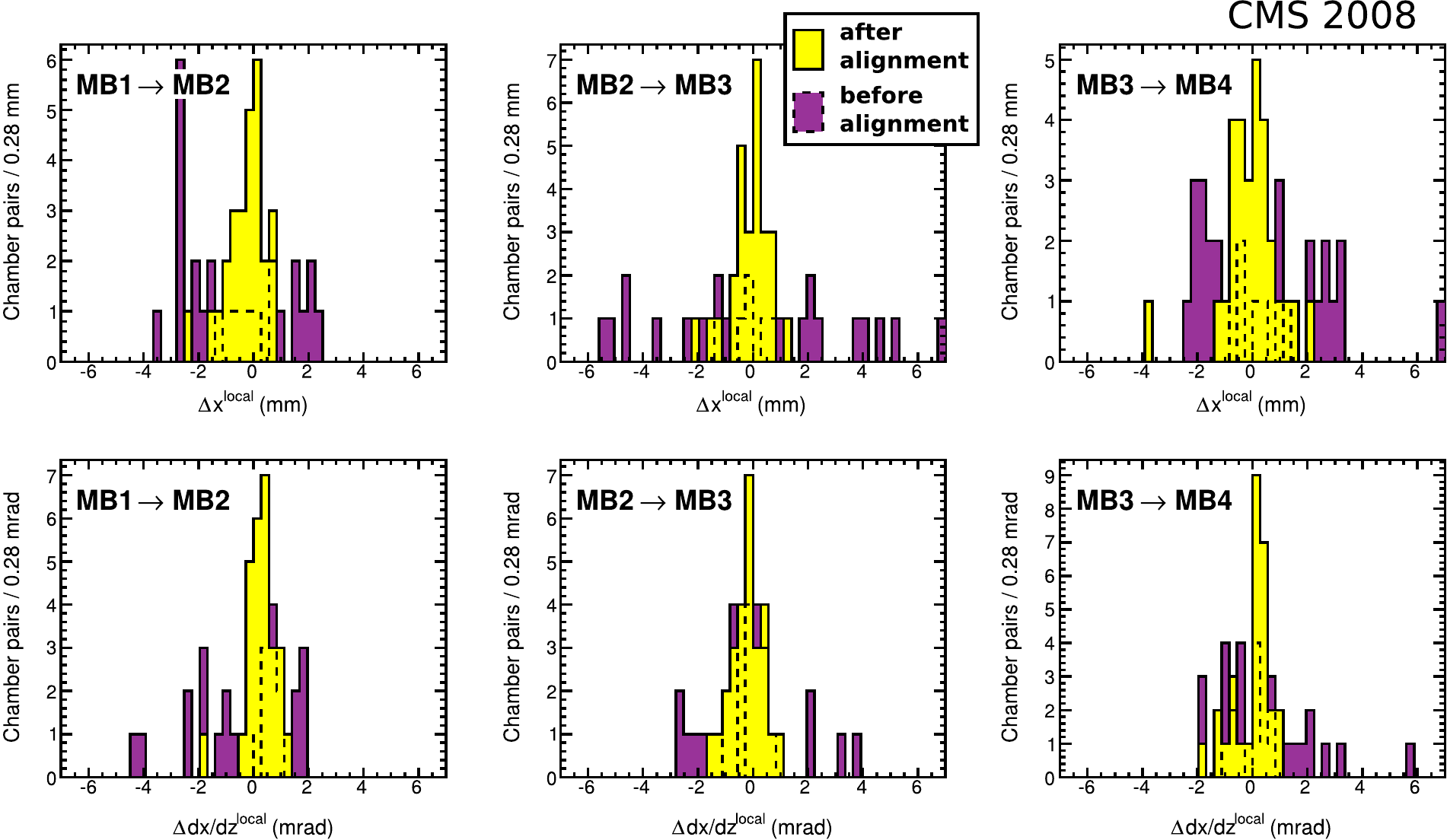}

  \caption{Differences in DT chamber positions and angles between pairs
  of stations as measured by locally propagated
  segments.  Dark/dashed histograms are before alignment; light histograms
  are after alignment.  \label{fig:valid_NOMvsMPvsHIP}}
\end{figure}

Figure~\ref{fig:valid_NOMvsMPvsHIP} shows the results of these
averages for each sector and pair of stations, before and after
alignment, and Table~\ref{tab:valid_NOMvsMPvsHIP} presents the RMS of
each of the presented distributions.  The distributions are convolutions of
errors in the alignment and errors in the segment-matching.  These results
therefore only set an upper limit on the
systematic uncertainties of the alignment itself.  In addition, global
distortions of the combined tracker and muon system are not quantified
by this method.  This results in an upper limit of 0.7~mm in
$\delta_x$ (proportional to $\Delta x$) and 0.6~mrad in
$\delta_{\phi_y}$ (approximately proportional to
$\Delta \frac{\textrm{d}x}{\textrm{d}z}$) for chambers in stations 1--3, and 1.0~mm,
0.7~mrad in station~4.

\begin{table}[p]
\caption{RMS of pairwise position and angle differences by station (distributions in Fig.~\ref{fig:valid_NOMvsMPvsHIP}). \label{tab:valid_NOMvsMPvsHIP}}
\centering

\vspace{0.25 cm}
\renewcommand{\arraystretch}{1.5}
\begin{tabular}{l | c c}
\hline\hline & \hspace{0.5 cm} Unaligned \hspace{0.5 cm} & \hspace{0.5 cm} Aligned \hspace{0.5 cm} \\\hline
MB1 $\to$ MB2 \hspace{0.25 cm} $\Delta x^{\mbox{\scriptsize local}}$ (mm) & 1.82 & 0.68 \\
MB1 $\to$ MB2 \hspace{0.25 cm} $\Delta \frac{\textrm{d}x}{\textrm{d}z}^{\mbox{\scriptsize local}}$ (mrad) & 1.68 & 0.57 \\\hline
MB2 $\to$ MB3 \hspace{0.25 cm} $\Delta x^{\mbox{\scriptsize local}}$ (mm) & 3.20 & 0.69 \\
MB2 $\to$ MB3 \hspace{0.25 cm} $\Delta \frac{\textrm{d}x}{\textrm{d}z}^{\mbox{\scriptsize local}}$ (mrad) & 1.56 & 0.60 \\\hline
MB3 $\to$ MB4 \hspace{0.25 cm} $\Delta x^{\mbox{\scriptsize local}}$ (mm) & 2.17 & 1.06 \\
MB3 $\to$ MB4 \hspace{0.25 cm} $\Delta \frac{\textrm{d}x}{\textrm{d}z}^{\mbox{\scriptsize local}}$ (mrad) & 1.65 & 0.70 \\\hline\hline
\end{tabular}
\end{table}

The uncertainty in point resolution along the line of sight of tracks
is therefore bounded between the Monte Carlo prediction
(Table~\ref{tab:hip_MC}), which includes only known propagation and
detector effects, and this diagnostic, which has its own systematic
uncertainties.  For stations~1--3, the uncertainty is at best 200 and
at worst 700~$\mu$m in $\delta_x$.

Muons from proton collisions, which illuminate the endcaps,
will enable a local cross-check with the CSC ring method of
Section~\ref{sec:localcsc}.  Since the latter has a demonstrated
position uncertainty of 270~$\mu$m, such a comparison will
considerably tighten the bound on alignment resolution uncertainty.

\subsection{Alignment Results: Effect on Momentum Measurement}
\label{sec:momentum}

The motivation for the alignment effort is to correct reconstructed
muon momentum distributions, so the trajectories of
cosmic-ray muons are re-fitted with the new geometry to verify that the resolution
has improved.  For sensitivity to the effect of misalignments in the muon
system, energetic cosmic rays are selected with $p_T > 200$~GeV/$c$, a
sample which is independent of the 100 $<$ $p_T$ $<$ 200~GeV/$c$
tracks used to perform the alignment.  Tracks are reconstructed using
hits from the tracker and the first muon station, a simple way to
optimize high-momentum muon resolution by increasing the effective
lever arm of sagitta measurements while minimizing bias from radiative
muon showers, which become prominent at several hundred GeV/$c$.  It
also focuses on the connection between the tracker and the first muon
station, a pair that was not tested with the segments described
in Section~\ref{sec:gmavalidation}.

\begin{figure}[p]
  \centering
  \includegraphics[width=0.48\linewidth]{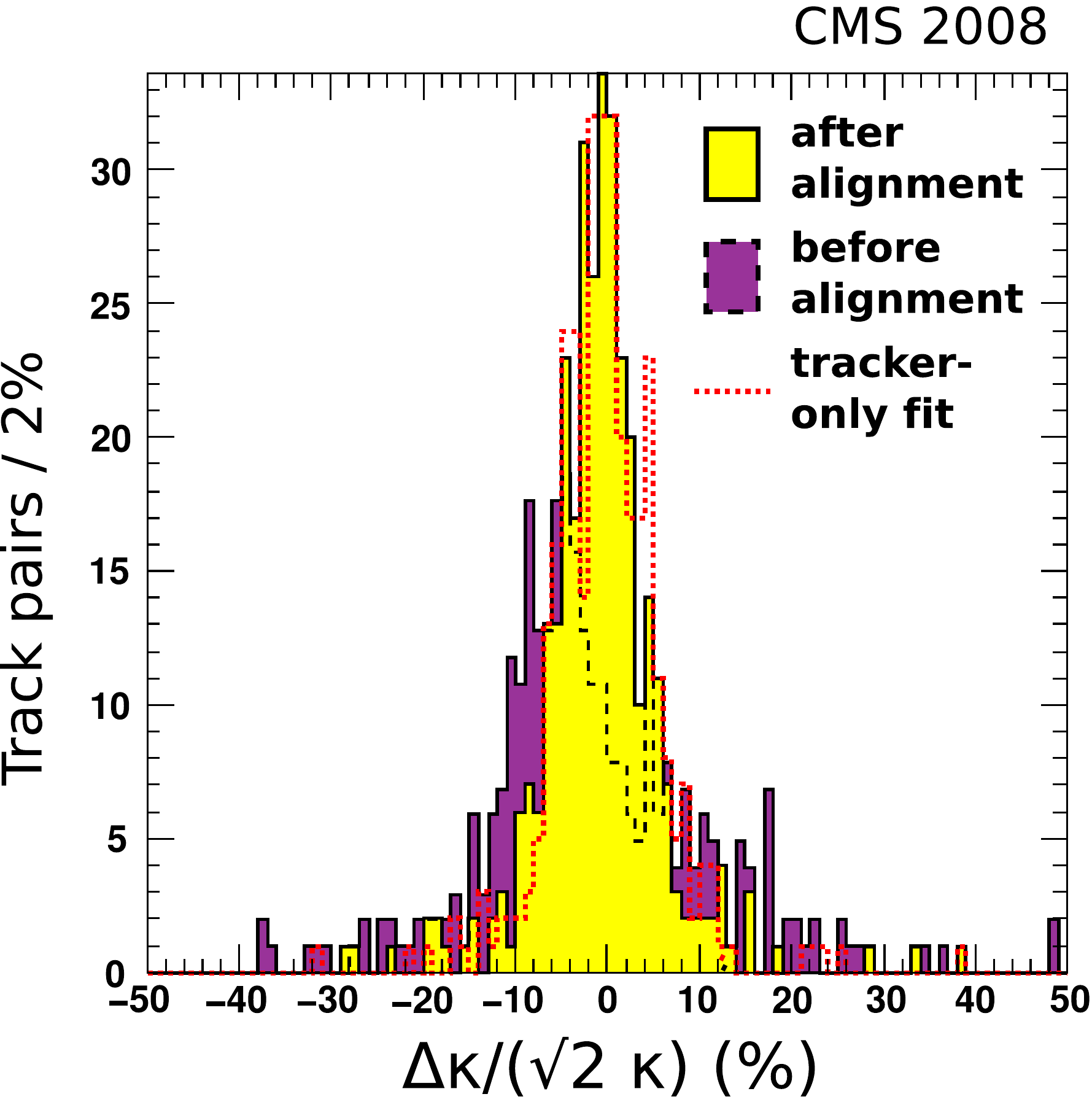}
  \caption{Fractional curvature difference between the top and bottom
  parts of CMS for muons with $p_T > 200$~GeV/$c$ ($\kappa = q/p_T$).
  Dark/dashed distribution is before alignment, light is after
  alignment (using the reference-target algorithm), and the open
  dotted distribution is tracker-only. \label{fig:chargesplitting}}
\end{figure}

The top half and bottom half of the cosmic ray trajectory are
reconstructed separately, split at the point of closest approach to
the LHC beamline.  Any difference in track parameters between the top
and bottom fits is purely instrumental: in
Fig.~\ref{fig:chargesplitting} the fractional difference in
curvature ($\Delta \kappa/(\sqrt{2}\kappa) =
(\kappa_{\mbox{\scriptsize top}} - \kappa_{\mbox{\scriptsize
bottom}})/(\sqrt{2}\kappa)$ where $\kappa = q/p_T$) is plotted before and after
alignment; the tracker-only reconstruction is also given, for
reference.  Assuming that the measurements in the top and bottom parts
of CMS are statistically independent with equal
resolutions, this plot represents the fractional error in the
curvature of tracks, which is approximately equal to the fractional
error in its reciprocal, $p_T$.  Some global distortions correlate
misalignments in the top and bottom parts of the detector; this test is
insensitive to such modes.  The fitted resolution is given in
Table~\ref{tab:chargesplitting}.

The majority of muons in this study
have a momentum close to the 200~GeV/$c$ threshold because of the
steeply falling distribution of cosmic rays.  At 200~GeV/$c$, the
momentum resolution of the combined system is not expected to
significantly exceed the resolution of the tracker alone (see
Ref.~\cite{Ball:2007zza}, Fig.~1.5), but the biased distribution prior
to alignment has been repaired by alignment procedure.

\begin{table}[p]
\caption{RMS and Gaussian core fits of the fractional curvature
  distributions in Fig.~\ref{fig:chargesplitting}.  The fits include
  all data in the central region
  $\left|\Delta \kappa/(\sqrt{2}\kappa)\right| <
  2\times$RMS. \label{tab:chargesplitting}}

\begin{center}
\renewcommand{\arraystretch}{1.5}
\begin{tabular}{l c c c}
\hline\hline & \hspace{0.25 cm}Gaussian mean (\%)\hspace{0.25 cm} & \hspace{0.25 cm}Gaussian width (\%)\hspace{0.25 cm} & \hspace{0.25 cm}RMS (\%)\hspace{0.25 cm} \\\hline
Unaligned & $-$2.5 $\pm$ 0.6 & 8.6 $\pm$ 0.5 & 12.3 \\
Aligned & $-$0.9 $\pm$ 0.3 & 4.3 $\pm$ 0.2 & 6.9 \\\hline
Tracker-only & \textcolor{white}{$-$}0.3 $\pm$ 0.3 & 4.5 $\pm$ 0.2 & 6.4 \\\hline\hline
\end{tabular}
\end{center}
\end{table}

\subsection{The Millepede Algorithm}
\label{sec:mpalgo}

A Millepede algorithm~\cite{Blobel:2006yh} is under development to
align the muon chambers relative to the tracker, as an alternative to
the reference-target algorithm.  Instead of using a general non-linear
fitting package to minimize the objective function, Millepede
linearizes the problem and solves it with matrix inversion.  A
potential application of this method is to combine in a single fit
measurements from tracker tracks and the locally fitted segments
described in Section~\ref{sec:gmavalidation}, thereby optimizing
statistical precision.  Before such a generalization is used, however,
it must be tuned to reproduce the results of the reference-target
algorithm.  When configured to use tracker tracks only, the objective
function is
\begin{equation}
\chi^2 = \sum_i^{\mbox{\scriptsize chambers}} \sum_i^{\mbox{\scriptsize tracks}}
\left(\Delta \vec{x}_j - A_j \cdot \vec{\delta}_i\right)^T
({\sigma_{\mbox{\scriptsize residual}}}^2)^{-1}_{ij}
\left(\Delta \vec{x}_j - A_j \cdot \vec{\delta}_i\right) \quad\mbox{,}
\label{eqn:gmamillepedechi2}
\end{equation}
where $\Delta \vec{x} = (\Delta x, \Delta
y, \Delta \frac{\textrm{d}x}{\textrm{d}z}, \Delta \frac{\textrm{d}y}{\textrm{d}z})$, $A_j$ is the
matrix in Eq.~(\ref{eqn:dtmatrix}), and
$({\sigma_{\mbox{\scriptsize residual}}}^2)^{-1}_{ij}$ is the inverse
of the residuals covariance matrix.

To avoid the influence of non-Gaussian tails, noisy channels, and
pattern-recognition errors in the determination of the peak of the residuals distribution,
large residuals were excluded, symmetrically around the peak, for
each of the four residuals components.  Threshold values for the
excluded region are derived from Cauchy-Lorentzian fits.  With $x_0$ and
$\gamma$ being the mean and half-width at half maximum,
only residuals between $x_0 - 2.5 \, \gamma$ and $x_0 + 2.5 \, \gamma$
enter the sum in Eq.~(\ref{eqn:gmamillepedechi2}) for all
four components.  The
optimum threshold was chosen from the study presented in
Table~\ref{tab:tailstudy}, which lists alignment position accuracy as
a function of the threshold value.

\begin{table}
\caption{Accuracy of Millepede alignment in Monte Carlo simulations ($350$~k tracks), varying the selection threshold of the
excluded region.  The resolution attained with a threshold of
$2.5 \, \gamma$ reproduces that of the reference-target
algorithm presented in
Table~\ref{tab:hip_MC}. \label{tab:tailstudy}}
\renewcommand{\arraystretch}{1.5}
\begin{center}
\begin{tabular}{c | c c c c c c}
\hline\hline selection threshold & $\delta_x$ (mm) & $\delta_y$ (mm) & $\delta_z$ (mm) \\\hline
$1.0 \, \gamma$ & 0.41 & 1.05 & 2.39 \\
$1.5 \, \gamma$ & 0.30 & 0.89 & 1.46 \\
$2.0 \, \gamma$ & 0.26 & 0.84 & 0.92 \\
$2.5 \, \gamma$ & 0.25 & 0.84 & 0.78 \\
$3.0 \, \gamma$ & 0.25 & 0.85 & 0.78 \\\hline\hline
\end{tabular}
\end{center}
\end{table}

\section{Summary and Discussion}

This paper has presented a variety of procedures to align
different parts of the muon system with tracks: layers in DT chambers,
CSC chambers in rings, and DT chambers in a coordinate frame shared
with the tracker.  The available data have been fully exploited:
horizontal beam-halo and vertical cosmic-ray muons.  Through
comparisons with independent data, it was shown that the superlayer
$r\phi$ resolution is 80~$\mu$m within each DT chamber, that the
$r\phi$ resolution of CSC chambers is 270~$\mu$m within each endcap
ring, and that the DT chamber positions along the tracks are known
with an accuracy between 200 and 700~$\mu$m (stations~1--3, wheels $-$1, 0,
$+$1).

In addition, several new techniques have been introduced.  The
superlayer structure of DT chambers permits an analysis of layer
geometry in a way that uses tracks alone, and therefore rigorously
compares the result obtained with tracks with the results from survey
measurements.  The overlap of CSC rings permits an analytic solution
to its alignment.  Non-Gaussianity in the physics of track propagation through the steel
yoke implies a non-linear extension to the general alignment method.

Techniques which will be useful for re-aligning
the muon system with early LHC data have been tested.  The favorable distribution of
muons from collisions will broaden the applicability of these methods
and open new opportunities for cross-checks and diagnostics, which
ultimately will lead to a better-understood momentum resolution for
high-momentum muons and increased discovery reach for high-energy
processes.

\section*{Acknowledgements}

We thank the technical and administrative staff at CERN and other CMS
Institutes, and acknowledge support from: FMSR (Austria); FNRS and FWO
(Belgium); CNPq, CAPES, FAPERJ, and FAPESP (Brazil); MES (Bulgaria);
CERN; CAS, MoST, and NSFC (China); COLCIENCIAS (Colombia); MSES
(Croatia); RPF (Cyprus); Academy of Sciences and NICPB (Estonia);
Academy of Finland, ME, and HIP (Finland); CEA and CNRS/IN2P3
(France); BMBF, DFG, and HGF (Germany); GSRT (Greece); OTKA and NKTH
(Hungary); DAE and DST (India); IPM (Iran); SFI (Ireland); INFN
(Italy); NRF (Korea); LAS (Lithuania); CINVESTAV, CONACYT, SEP, and
UASLP-FAI (Mexico); PAEC (Pakistan); SCSR (Poland); FCT (Portugal);
JINR (Armenia, Belarus, Georgia, Ukraine, Uzbekistan); MST and MAE
(Russia); MSTDS (Serbia); MICINN and CPAN (Spain); Swiss Funding
Agencies (Switzerland); NSC (Taipei); TUBITAK and TAEK (Turkey); STFC
(United Kingdom); DOE and NSF (USA). Individuals have received support
from the Marie-Curie IEF program (European Union); the Leventis
Foundation; the A. P. Sloan Foundation; and the Alexander von Humboldt
Foundation.

%---------------------------------------------------------------------
\bibliography{auto_generated}   % will be created by the tdr script.
\cleardoublepage\appendix\section{The CMS Collaboration \label{app:collab}}\begin{sloppypar}\hyphenpenalty=500\textbf{Yerevan Physics Institute,  Yerevan,  Armenia}\\*[0pt]
S.~Chatrchyan, V.~Khachatryan, A.M.~Sirunyan
\vskip\cmsinstskip
\textbf{Institut f\"{u}r Hochenergiephysik der OeAW,  Wien,  Austria}\\*[0pt]
W.~Adam, B.~Arnold, H.~Bergauer, T.~Bergauer, M.~Dragicevic, M.~Eichberger, J.~Er\"{o}
% ...
\vskip\cmsinstskip
% ...

\end{sloppypar}
\end{document}